\documentclass[conference,10pt,letterpaper]{IEEEtran}

\usepackage[colorlinks,citecolor=blue]{hyperref}

\usepackage{booktabs}

\usepackage[caption=false]{subfig}

\usepackage{multirow}

\usepackage[pdftex]{graphicx}
\graphicspath{{figures/measurement_error/}{figures/pb_training/}{figures/overlap/}{figures/privacy_protection/world_1/}{figures/privacy_protection/world_2/}{figures/website_quality_metrics/}{figures/website_quality_manual/}{figures/synthetic/}}

\begin{document}
	
\title{A comparison of web privacy protection techniques}

\author{\IEEEauthorblockN{Johan Mazel}
\IEEEauthorblockA{National Institute of Informatics/JFLI\\
johan.mazel@ssi.gouv.fr}
\and
\IEEEauthorblockN{Richard Garnier}
\IEEEauthorblockA{ENSIMAG\\
  garnier.richard.s@gmail.com}
\and
\IEEEauthorblockN{Kensuke Fukuda}
\IEEEauthorblockA{National Institute of Informatics/Sokendai\\
kensuke@nii.ac.jp}}

\IEEEoverridecommandlockouts
\makeatletter\def\@IEEEpubidpullup{9\baselineskip}\makeatother
\IEEEpubid{\parbox{\columnwidth}{}
	\hspace{\columnsep}\makebox[\columnwidth]{}}

\maketitle

\begin{abstract}
Tracking is pervasive on the web.
Third party trackers acquire user data through information leak from websites, and user browsing history using cookies and device fingerprinting.
In response, several privacy protection techniques (e.g. the Ghostery browser extension) have been developed.
To the best of our knowledge, our work is the first study that proposes a reliable methodology for privacy protection comparison, and extensively compares a wide set of privacy protection techniques.
Our contributions are the following.
First, we propose a robust methodology to compare privacy protection techniques when crawling many websites, and quantify measurement error.
To this end, we reuse the privacy footprint \cite{Krishnamurthy2006Generating} and apply the Kolmogorov-Smirnov test on browsing metrics.
This test is likewise applied to HTML-based metrics to assess webpage quality degradation.
To complement HTML-based metrics, we also design a manual analysis.
Second, we study the overlap of blocking resources between most popular browser extensions, and compare the performances using the proposed methodology.
We show that protection techniques have vastly different performances, and that the best of them exhibit a wide overlap.
Next, we analyze the impact of privacy protection techniques on webpage quality. 
We show that automated HTML-based analysis sometimes fails to expose quality reduction perceived by users. 
Finally, we provide a set of usage recommendations for end-users and research recommendations for the scientific community.
Ghostery and uBlock Origin provide the best trade-off between protection and webpage quality.
Ghostery however requires a configuration step which is difficult for users.
The RequestPolicy Continued and NoScript extensions exhibit the best performances but reduce webpage quality.
Ghostery and uBlock Origin use manually built blocking lists which are cumbersome to maintain.
Research efforts should focus on improving existing approaches that do not rely on blocking lists (such as Privacy badger or MyTrackingChoices), and automatically building reliable blocking lists.  
\end{abstract}

\newcommand{\adb}{AdBlock Plus }
\newcommand{\blur}{Blur }
\newcommand{\dnt}{DoNotTrack }
\newcommand{\gho}{Ghostery }
\newcommand{\ns}{NoScript }
\newcommand{\nt}{NoTrace }
\newcommand{\pb}{Privacy Badger }
\newcommand{\dec}{Decentraleyes }
\newcommand{\rpc}{RequestPolicy Continued }
\newcommand{\dis}{Disconnect }
\newcommand{\ubo}{uBlock Origin }
\newcommand{\el}{EasyList }
\newcommand{\ep}{EasyPrivacy }
\newcommand{\mtc}{MyTrackingChoices }
\newcommand{\bt}{BeefTaco }
\newcommand{\https}{HTTPSEverywhere }
\newcommand{\wot}{WebOfTrust }

\section{Introduction}
The huge growth of the Internet comes along with an ever-increasing advertising market.
Internet users access content provided for free by publishers.
Consequently, publishers monetize their audience through advertisement.
Companies thus buy online exposure to promote their products.
In order to maximize advertisement efficiency, advertisers tailor ads to users regarding their interests.
To this end, advertisers leverage context (e.g. visited website) or previous browsing interests.

Advertisers use techniques such as cookies to identify users across websites and build their browsing history.
Other techniques have also been developed to allow advertising actors to communicate with each other (such as cookie syncing~\cite{Acar2014The}), or circumvent cookie removal by respawning cookies using diverse types of data storage inside the browser (e.g. using Flash~\cite{Soltani2010Flash}).
Browser fingerprinting~\cite{Eckersley2010How} allows a tracking entity to follow a user across websites without any in-browser data storage.
In response to these techniques, several counter-measures were designed.
We can here quote the Do Not Track HTTP header~\cite{DoNotTrack} by which a user can ask not to be tracked.
Browsers can also block some or all cookies.
Finally, many browser extensions hinder third party tracking by preventing cookie creation and/or blocking requests to tracking services.

End-users thus have many techniques available but have trouble picking one that offers good protection.
Similarly, privacy researchers often want to assess protection efficiency but do not want to test all available tools.
Our goal is to compare existing privacy protection techniques and provide efficiency-based recommendations.
Some work previously tried to compare privacy protection techniques \cite{Krishnamurthy2007Measuring,Mayer2012Third,Balebako2012Measuring,Hill2014Comparative,Hill2015uBlock,Wills2016What,Merzdovnik2017Block,Traverso2017Benchmark}. 
Krishnamurthy et al.~\cite{Krishnamurthy2007Measuring} provide the first comparison of privacy protection methods but do not evaluate most of the browser extensions available today because they appeared after the publication of the paper.
Balebako et al.~\cite{Balebako2012Measuring} focus on a very specific use case: behavioral advertising.
Mayer et al.~\cite{Mayer2012Third}, Hill~\cite{Hill2014Comparative,Hill2015uBlock} and Traverso et al.~\cite{Traverso2017Benchmark} evaluate 4, 4, 4 and 7 browser extensions on a limited set of websites (respectively Alexa Top 500, 45, 84 and 100 italian URLs).
Wills et al.~\cite{Wills2016What} crawl a thousand of websites but use only six browser extensions.
Merzdovnik et al.~\cite{Merzdovnik2017Block} use five browser extensions.
Furthermore, these studies do not use the same metrics, it is thus difficult to compare their results.
Our extensive coverage improves the state of the art regarding measurement methodology and error, and evaluated techniques.

Our contributions are the following.
First, we propose a robust methodology to compare privacy protection techniques against third party tracking when crawling many websites.
To this end, we reuse the privacy footprint \cite{Krishnamurthy2006Generating} and apply the Kolmogorov-Smirnov test on browsing metrics.
This test is likewise applied to HTML-based metrics to assess webpage quality degradation.
We also design a manual analysis to complement HTML-based metrics.
Second, we analyze the blocked resource overlap between privacy protection techniques, and compare their performances.
Third, we assess the impact of privacy protection techniques on website quality.
Finally, we provide recommendations for end-users and scientific community.

\section{Background}
\label{sec:background}

This section presents an overview of third party web tracking and existing privacy protection techniques.
We refer the reader to \cite{Bujlow2017Web,Estrada2016Online} for a more complete description of these aspects.

\subsection{Third party web tracking}
\label{sec:web_tracking}

Websites massively rely on advertisement to monetize user visit.
Advertisers purchase ads for their products directly from publishers or ad exchanges.  
When users access websites, they communicate with all these actors.
From the users' viewpoint, these entities belong to two categories: first and third parties.
First parties are the entities that users intend to reach, here, publishers.
Third parties can be advertisers, ad exchanges or others actors that provide services to first parties (such as web analytics).
Using techniques such as cookies, local data storage or fingerprinting, third parties can identify users across websites. 
Combining HTTP referrer field and user identification, they can reconstruct users' browsing history.
Using cookie syncing, trackers can also exchange user identification and thus improve data collection.

\newcommand{\cm}{\ding{51}}
\newcommand{\no}{\ }

\begin{table}
  \centering
  \setlength\tabcolsep{3pt} %
  \begin{tabular}{cllrr}
    \toprule
    
     & \textbf{Name} & \textbf{License} & \textbf{User number (K)} & \textbf{Version} \\
    
    \midrule
        
    \multirow{7}{*}{\rotatebox{90}{\textbf{Blocking lists}}}
    
     & AdBlock Plus \cite{AdBlockPlus}                   & GPL v3      & 13,917 &      2.7 \\
     & uBlock Origin \cite{uBlockOrigin}                 & GPL v3      &  3,759 &      1.6 \\
     & Ghostery \cite{Ghostery}                          & Proprietary &    966 &   5.4.10 \\
     & Disconnect \cite{Disconnect}                      & GPL v3      &    204 &   3.15.3 \\
     & NoTrace \cite{NoTrace}                            & -           &    0.8 &      2.4 \\
     & DoNotTrackMe/Blur \cite{Blur}                     & Prop.       &     86 & 6.0.2091 \\
     & BeefTaco \cite{Achara2016My}                      & Apa. 2.0    &     10 &  1.3.7.1 \\
     
     \midrule
     \multirow{2}{*}{\rotatebox{90}{\textbf{Heur.}}}
     & Privacy Badger \cite{PrivacyBadger}               & GPL v3      &    205 &    1.0.6 \\
     & MyTrackingChoices \cite{Achara2016My}             & -           &    0.1 &      1.0 \\
     
     \midrule
     \multirow{2}{*}{\rotatebox{90}{\textbf{Ind.}}}
     & NoScript \cite{NoScript}                          & GPL v2      &  1,765 &      2.9 \\
     & RPC \cite{RequestPolicyContinued}                 & GPL v3      &      6 &      1.0 \\
     
     \midrule
     \multirow{3}{*}{\rotatebox{90}{\textbf{Other}}}
     & HTTPSEverywhere \cite{HTTPSEverywhere}            & GPL v2      &    353 &    5.2.5 \\
     & Decentraleyes \cite{Decentraleyes}                & MPL 2.0     &     69 &    1.2.2 \\
     & WebOfTrust \cite{WebOfTrust}                      & GPL v3      &    315 &        - \\
    \bottomrule
    
  \end{tabular}

  \vspace{5mm}
  
  \caption{Privacy protection technique characteristics. Heur.: heuristics ; Ind.: Indiscriminate ; Popularity (K): popularity in thousands.}
  \label{table:protection_techniques}
\end{table}

\subsection{Privacy protection techniques}
\label{sec:protection_techniques}

Several techniques have been designed to protect users from third party tracking.
Network-based techniques use DNS filtering or proxy.
They however exhibit several shortcomings: proxies cannot analyze encrypted traffic while DNS filtering only blocks entire domains \cite{Merzdovnik2017Block}.
User agent spoofing browser extensions may also improve privacy by hindering fingerprinting, but exhibit poor performance in practice \cite{Nikiforakis2013Cookieless}.
The next subsections provide a breakdown of browser-related techniques that we compare in this work.
Unless specified otherwise, these extensions are open source.

\subsubsection{Extensions}
\label{sec:protection_extensions}

We classify extensions regarding used third party tracking impediment methods: blocking lists, heuristics, indiscriminate blocking, or other.

The following browser extensions use blocking lists made of regex-based rules on domain names.
These lists are usually community maintained.
\gho \cite{Ghostery} is a proprietary, privacy-focused extension that uses a specific tracker blocking list.
It has recently been bought by the privacy-focused browser Cliqz \cite{Vincent2017Ghostery}.
Ghostery requires a configuration step to select categories of trackers to block.
\ubo \cite{uBlockOrigin} is a general purpose blocker. 
It can also understand the syntax used by the famous ad-blocker AdBlock Plus.
\ubo includes by default: EasyList, EasyPrivacy, Peter Lowe's Adservers, Malware domains and some specific lists.
\dis \cite{Disconnect} is another blocker.
\blur \cite{Blur} (also known as Do Not Track Me) is a proprietary extension owned by Abine.
It blocks trackers, protects mail addresses and passwords.
\nt \cite{NoTrace} uses a wide range of techniques in order to enhance privacy on the Internet.
NoTrace needs to be configured after installation.
One can also block tracking by setting opt-out cookies that block domains from setting standard cookies.
Several entities \cite{NAI,DAA} provide opt-cookie lists.
\bt \cite{BeefTaco} creates opt-out cookies in the browser.
Another privacy protection approach uses ad-blockers to hinder tracking in the same way they block advertisements.
\adb \cite{AdBlockPlus} is the most popular ad blocker.
Using specific lists, it can block trackers, social widgets, or malwares. 
Since 2011, \adb is commercially exploited by Eyeo which monetizes domain whitelisting
\cite{AcceptableAds}.
In addition to the ad-focused \el \cite{EasyList_EasyPrivacy} that is used by default in AdBlock Plus, there is a privacy-focused list called EasyPrivacy \cite{EasyList_EasyPrivacy}.

Unlike previous extensions relying on blocking list, some use heuristics to block trackers.
\pb \cite{PrivacyBadger} is developed by the Electronic Frontier Foundation and its behavior is further described in \autoref{sec:pb_training}.
By default, Privacy Badger uses the Do Not Track HTTP header~\cite{DoNotTrack} and strips the referrer field in HTTP requests.
\mtc \cite{Achara2016My} is an advertisement friendly privacy protection extension.
It uses an hybrid approach which leverages both heuristics similar to \pb and blocking list for bootstrapping.
Users can allow tracking for some website categories.

Some extensions indiscriminately block resources
\ns \cite{NoScript} disables JavaScript.
As a side effect, this also disables some tracking.
\rpc \cite{RequestPolicyContinued} blocks all third party requests.

Finally, some extensions use other mechanisms to protect users' privacy.
\https \cite{HTTPSEverywhere} tries to replace HTTP connections with HTTPS ones if HTTPS is supported.
\wot (WOT) \cite{WebOfTrust} provides website rating regarding trustworthiness and child safety.
It was temporarly removed from extensions stores when it was revealed that WOT broke privacy rules of browser developers 
\cite{Eckert2016Web}.
\dec uses local files to emulate resources, e.g. freely available JavaScript libraries, hosted on centralized entities.
This blocks tracking from these entities.
\autoref{table:protection_techniques} provides popularity (measured as the number of users for Firefox), source code license and version of the extensions used in our work.

\subsubsection{Browsers}

Some browsers also have built-in privacy protection features.
\dnt \cite{DoNotTrack} is a field in the HTTP header that asks the destination not to track the sender.
It was proposed in 2009 and is currently under standardization in W3C.
This feature is turned off by default in Firefox.
The last proposed amendments to the European Union's ePrivacy Regulation hints at an increase interest of policy makers towards \dnt \cite{Olejnik2017Proposed}.
Firefox \cite{Firefox} can block cookies: either all or only third parties' ones or only third party cookies from previously visited domains.
A tracking protection \cite{Kontaxis2015Tracking} has also been added in Firefox version 42.
Brave \cite{Brave} and Cliqz \cite{CLiqz} are browsers that emphasize their privacy protection features.

\newcommand\ExtraSep
{\dimexpr\cmidrulewidth+\aboverulesep+\belowrulesep\relax}

\newcommand{\us}{$\cdot$}
\newcommand{\gr}{$\bigcirc$}
\newcommand{\gu}{$\odot$}

\newcommand{\co}{$+$}
\newcommand{\cg}{$\oplus$}
\newcommand{\ml}{$\star$}

\newcommand{\re}{$\sim$}

\begin{table*}[ht!]
    \centering
    \setlength\tabcolsep{3pt} 
    \renewcommand{\arraystretch}{0.5}
    
    \resizebox{\textwidth}{!}{%
    \begin{tabular}{llccccccccccccccccccc}
        
        \toprule
        \textbf{Type} & \textbf{Authors \& references} & \multicolumn{17}{c}{\textbf{Extensions}} & \multicolumn{2}{c}{\textbf{Browsers}} \\
        
        \cmidrule(lr){3-19} \cmidrule(lr){20-21}
        
        & & 
        \rotatebox{90}{\textbf{AdBlock \cite{AdBlock}}} & 
        \rotatebox{90}{\textbf{AdBlock Plus EL \cite{AdBlockPlus,EasyList_EasyPrivacy}}} & 
        \rotatebox{90}{\textbf{AdBlock Plus EP \cite{AdBlockPlus,EasyList_EasyPrivacy}}} & 
        \rotatebox{90}{\textbf{uBlockOrigin \cite{uBlockOrigin}}} & 
        \rotatebox{90}{\textbf{Ghostery \cite{Ghostery}}} & 
        \rotatebox{90}{\textbf{Privacy Badger \cite{PrivacyBadger}}} & 
        \rotatebox{90}{\textbf{Disconnect \cite{Disconnect}}} & 
        \rotatebox{90}{\textbf{NoTrace \cite{NoTrace}}} & 
        \rotatebox{90}{\textbf{WebOfTrust \cite{WebOfTrust}}} & 
        \rotatebox{90}{\textbf{DoNotTrackMe/Blur \cite{Blur}}} & 
        \rotatebox{90}{\textbf{MyTrackingChoices \cite{Achara2016My}}} & 
        \rotatebox{90}{\textbf{RPC \cite{RequestPolicyContinued}}} & 
        \rotatebox{90}{\textbf{Decentraleyes \cite{Decentraleyes}}} & 
        \rotatebox{90}{\textbf{NoScript \cite{NoScript}}} & 
        \rotatebox{90}{\textbf{DNT \cite{DoNotTrack}}} & 
        \rotatebox{90}{\textbf{Opt-out cookies \cite{NAI,DAA,BeefTaco}}} & 
        \rotatebox{90}{\textbf{HTTPSEverywhere \cite{HTTPSEverywhere}}}
        & \rotatebox{90}{\textbf{Firefox \cite{Kontaxis2015Tracking}}} & 
        \rotatebox{90}{\textbf{Cliqz}} \\
        \midrule
        
        & Krishnamurthy et al \cite{Krishnamurthy2006Generating} &     & \gr &     &     &     &     &     &     &     &     &     &     &     &     &     &     &     &     &     \\    
        & Castelluccia et al \cite{Castelluccia2013Data}         &     & \gr &     &     & \gr &     &     &     &     &     &     &     &     &     &     &     &     &     &     \\    
        Web
        & Fruchter et al \cite{Fruchter2015Variations}           &     & \gr & \gr &     &     &     &     &     &     &     &     &     &     &     &     &     &     &     &     \\ 
        privacy
        & Carrascosa et al \cite{Carrascosa2015Always}           &     &     &     &     &     &     &     &     &     &     &     &     &     &     & \co &     &     &     &     \\   
        analysis
        & Metwalley et al \cite{Metwalley2015The}                & \us & \us &     &     & \gu &     &     &     & \us & \gu &     &     &     & \us &     &     &     &     &     \\                              
        & Englehardt et al \cite{Englehardt2015Cookies}           &     &     &     &     & \co &     &     &     &     &     &     &     &     &     & \co &     & \co & \co &     \\                              
        & Englehardt et al \cite{Englehardt2016Online}            &     &\gr\co&\gr\co&   & \co &     &     &     &     &     &     &     &     &     &     &     &     & \co &     \\    
        \midrule

        & Malandrino et al \cite{Malandrino2013PrivacyAwareness} &     & \co &     &     & \co &     &     & \co &     &     &     & \co &     & \co &     &     &     &     &     \\    
        Privacy
        & Malandrino et al \cite{Malandrino2013PrivacyLeakage}   &     & \co &     &     & \co &     &     & \co &     &     &     & \co &     & \co &     &     &     &     &     \\    
        protection
        & Kontaxis et al \cite{Kontaxis2015Tracking}             &     & \co &     &     &     &     &     &     &     &     &     &     &     &     &     &     &     & \co &     \\    
        & Achara et al \cite{Achara2016My}                       &     & \co &     & \co & \co &     &     &     &     &     & \co &     &     &     &     &     &     & \co &     \\
        \midrule

        & Gugelmann et al \cite{Gugelmann2015Automated}          &     &\co\ml&\co\ml&   &     &     &     &     &     &     &     &     &     &     &     &     &     &     &     \\
        Tracker
        & Metwalley et al \cite{Metwalley2015Unsupervised}       &     &     &     &     & \gr &     &     &     &     & \gr &     &     &     &     &     &     &     &     &     \\    
        detection
        & Wu et al \cite{Wu2015Tracker}                           &     &     &     &     & \ml &     &     &     &     &     &     &     &     &     &     &     &     &     &     \\    
        & Yu et al \cite{Yu2016Tracking}                         &     &     &     &     &     &     & \co &     &     &     &     &     &     &     &     &     &     & \co & \co \\
        & Ikram et al \cite{Ikram2017Towards}                    &     & \co &     &     & \co & \co & \co &     &     &     &     &     &     & \co &     &     &     &     &     \\
        \midrule

        User
        & Leon et al \cite{Leon2012Why}                          &     & \co &     &     & \co &     &     &     &     &     &     &     &     &     & \co & \co &     &     &     \\    
        ana.
        & Malandrino et al \cite{Malandrino2013How}              &     &     &     &     &     &     &     & \co &     &     &     &     &     &     &     &     &     &     &     \\
        \midrule

        \multirow{7}{*}{Comp.}
        & Krishnamurthy et al \cite{Krishnamurthy2007Measuring}  &     & \co &     &     &     &     &     &     &     &     &     &     &     & \co &     &     &     &     &     \\    
        & Mayer et al \cite{Mayer2012Third}                      &     & \co & \co &     & \co &     &     &     &     &     &     &     &     &     &     & \co &     &     &     \\    
        & Balebako et al \cite{Balebako2012Measuring}            &     & \co &     &     & \co &     &     &     &     &     &     &     &     &     & \co & \co &     &     &     \\    
        & Hill \cite{Hill2014Comparative}                         &     & \co &     &     & \co & \co & \co &     &     &     &     &     &     &     &     &     &     &     &     \\    
        & Hill \cite{Hill2015uBlock}                              &     & \co &     & \co & \co &     & \co &     &     &     &     &     &     &     &     &     &     &     &     \\
        & Wills et al \cite{Wills2016What}                       &     & \co & \co & \co & \co &     & \co &     &     & \co &     &     &     &     &     &     &     &     &     \\
        & Merzdovnik et al \cite{Merzdovnik2017Block}            &     & \co &     & \co & \co & \co & \co &     &     &     &     &     &     &     &     &     &     &     &     \\
        & Traverso et al \cite{Traverso2017Benchmark}            &     & \co & \gr & \co &\gr\co& \co&\gr\co&    &     & \co &     & \co &     &     &     &     &     &     &     \\
        \midrule
        \midrule
        Our work & -                                              & \re & \co & \co & \co & \co & \co & \co & \co & \co & \co & \co & \co & \co & \co & \co & \co & \co & \re &     \\
        \bottomrule    
        
    \end{tabular}
    }

    \caption{Extensions used in previous publications (\us\ for usage in the wild, \gr\ for ground-truth, \gu\ for ground-truth and usage, \co\ for comparison, \cg\ for comparison and ground-truth, \ml\ for ML bootstrapping , \re means that we did not evaluate these techniques but that their results can be derived from our work: AdBlock uses the same default blocking list as AdBlock Plus and Firefox Tracking Protection uses the same blocking list as Disconnect).}
    \label{table:related_work_techniques}
    
\end{table*}

\section{Related work}
\label{sec:related_work}

\subsection{Tracking on the Internet}

The seminal contribution in the field of web privacy was by Krishnamurthy et al.
They first studied the diffusion of privacy information \cite{Krishnamurthy2006Generating}, then, analyzed the impact of counter-measures on this diffusion and webpage quality \cite{Krishnamurthy2007Measuring}, and observed the increase of user-data aggregation by a small number of entities \cite{Krishnamurthy2009Privacy}.
Several works then tried to analyze this phenomenon with more details by classifying tracking \cite{Roesner2012Detecting} or analyzing geographical variations of tracking \cite{Castelluccia2013Data,Fruchter2015Variations}.

While early studies focused on cookie-based tracking, two main new tracking techniques trends emerged: resilient in-browser data storage \cite{Soltani2010Flash} and browser fingerprinting \cite{Eckersley2010How}.
Local data storage allows one to store data in multiple locations (for example, Flash Local Storage Object, HTML5 or ETags among other means) on a device to bypass standard cookie removal.
This technique thus provides resilient local identifier storage.
By using browser fingerprint, a tracker can completely avoid using and storing an identifier inside the browser and instead recognize a browser, or the device it is installed on, using its characteristics such as fonts \cite{Fifield2015Fingerprinting}, battery \cite{Olejnik2015Leaking}, canvas \cite{Mowery2012Pixel} or hardware level features \cite{Cao2017Cross}.
Several studies then tried to detect fingerprinting \cite{Acar2013FP}, or both local  storage and fingerprinting \cite{Acar2014The,Englehardt2016Online}.
Furthermore, Lerner et al. \cite{Lerner2016Internet} show that local storage fingerprinting increased between 1996 and 2016.
Another privacy threat is the practice of sharing user identification across tracking entities. 
This is called ID-sharing, cookie-syncing or cookie-matching.
While the phenomenon has been documented for some time \cite{Ghosh2011Selling}, recent work analyzed its prevalence in the wild and its impact on user browsing history sharing among trackers \cite{Acar2014The,Olejnik2014Selling,Englehardt2016Online}.

Finally, the feasibility of user surveillance through bulk passive network traffic monitoring-based cookie observation has also been analyzed \cite{Englehardt2015Cookies}.

\newcommand{\cd}{$\bigcirc$}
\newcommand{\cb}{$\odot$}

\begin{table*}[t!]
    \centering
    \setlength\tabcolsep{3pt}
    \renewcommand{\arraystretch}{0.5}
    
    \resizebox{\textwidth}{!}{%
    \begin{tabular}{llccccccccccccccccccccc}
        
        \toprule
        \textbf{Type} & 
        \textbf{Authors \& references} & 
        \multicolumn{10}{c}{\textbf{Privacy protection}} &
        \multicolumn{6}{c}{\textbf{Webpage quality}} & 
        \multicolumn{3}{c}{\textbf{Browser}} & 
        & \\
        
        \cmidrule(lr){3-12} \cmidrule(lr){13-18} \cmidrule(lr){19-21}
        
        &  & 
        \rotatebox{90}{\textbf{\# domains}} & 
        \rotatebox{90}{\textbf{\# 3rd party domains}} & 
        \rotatebox{90}{\textbf{\# hosts}} & 
        \rotatebox{90}{\textbf{\# cookies}} & 
        \rotatebox{90}{\textbf{\# third party cookies}} & 
        \rotatebox{90}{\textbf{\# requests}} & 
        \rotatebox{90}{\textbf{\# first party requests}} & 
        \rotatebox{90}{\textbf{\# third party requests}} & 
        \rotatebox{90}{\textbf{TP/FP/FN/TN}} & 
        \rotatebox{90}{\textbf{Privacy footprint}} & 
        \rotatebox{90}{\textbf{\# script}} & 
        \rotatebox{90}{\textbf{\# image}} & 
        \rotatebox{90}{\textbf{Data}} & 
        \rotatebox{90}{\textbf{Script data size}} & 
        \rotatebox{90}{\textbf{Image data size}} & 
        \rotatebox{90}{\textbf{Manual analysis}} & 
        \rotatebox{90}{\textbf{Loading time}} & 
        \rotatebox{90}{\textbf{Browser CPU usage}} & 
        \rotatebox{90}{\textbf{Browser memory usage}} & 
        \rotatebox{90}{\textbf{Use-case specific metric}}
        \\
        \midrule
        
        & Krishnamurthy et al. \cite{Krishnamurthy2006Generating} &     &     &     &     &     &     &     &     &     & \cd &     &     &     &     &     &     &     &     &     &     &     \\
        Web
        & Fruchter et al. \cite{Fruchter2015Variations}           &     &     &     &     & \cd &     &     & \cd &     &     &     &     &     &     &     &     &     &     &     &     &     \\
        privacy
        & Carrascosa et al. \cite{Carrascosa2015Always}           &     &     &     &     &     &     &     &     &     &     &     &     &     &     &     &     &     &     &     & \cd &     \\
        analysis
        & Englehardt et al. \cite{Englehardt2015Cookies}          &     &     &     &     &     &     &     &     &     &     &     &     &     &     &     &     &     &     &     & \cd &     \\
        & Englehardt et al. \cite{Englehardt2016Online}           &     &     &     &     & \cd &     &     & \cd &     &     & \cd &     &     &     &     &     &     &     &     & \cd &     \\
        \midrule
        
        & Malandrino et al. \cite{Malandrino2013PrivacyAwareness} &     &     &     &     &     &     &     &     & \cd &     &     &     &     &     &     &     & \cd &     & \cd &     &     \\
        Privacy
        & Malandrino et al. \cite{Malandrino2013PrivacyLeakage}   &     &     &     &     &     &     &     &     &     &     &     &     & \cb &     &     &     &     &     &     &     &     \\
        protection
        & Kontaxis et al. \cite{Kontaxis2015Tracking}             &     &     &     &     &     &     &     & \cd &     &     &     &     & \cd &     &     &     & \cd &     &     &     &     \\
        & Achara et al. \cite{Achara2016My}                       &     &     &     &     &     &     &     &     &     &     &     &     &     &     &     &     & \cd & \cd & \cd &     &     \\
        \midrule
        
        Tracker
        & Yu et al. \cite{Yu2016Tracking}                         &     &     &     &     &     &     &     & \cd &     &     &     &     &     &     &     &     &     &     &     &     &     \\
        detection
        & Ikram et al. \cite{Ikram2017Towards}                    &     &     &     &     &     &     &     &     &     &     & \cd &     &     &     &     &     &     &     &     &     &     \\
        \midrule
        
        \multirow{7}{*}{Comparison} 
        & Krishnamurthy et al. \cite{Krishnamurthy2007Measuring}  & \cb &     &     &     &     &     &     &     &     &     &     &     & \cd &     &     &     &     &     &     &     &     \\
        & Mayer et al. \cite{Mayer2012Third}                      &     &     &     & \cd &     & \cd &     &     &     &     &     &     &     &     &     &     &     &     &     &     &     \\
        & Balebako et al. \cite{Balebako2012Measuring}            &     &     &     & \cd &     &     &     &     &     &     &     &     &     &     &     &     &     &     &     & \cd &     \\
        & Hill \cite{Hill2014Comparative}                         & \cd &     & \cd & \cd &     & \cd &     &     &     &     & \cd &     & \cd &     &     &     &     &     &     &     &     \\
        & Hill \cite{Hill2015uBlock}                              &     &     &     & \cd &     & \cd &     &     &     &     & \cd &     &     &     &     &     &     & \cd &     &     &     \\
        & Wills et al. \cite{Wills2016What}                       &     & \cd &     &     &     &     &     &     &     &     &     &     &     &     &     &     &     &     &     &     &     \\
        & Merzdovnik et al. \cite{Merzdovnik2017Block}            &     &     &     &     &     &     &     & \cd &     &     &     &     &     &     &     &     &     &     &     &     &     \\
        & Traverso et al. \cite{Traverso2017Benchmark}            &     & \cb &     &     &     &     &     &     &     &     &     &     & \cd &     &     &     & \cd &     &     &     &     \\
        \midrule
        \midrule
        Our work & -                                              &     & \cd &     & \cd &     & \cb & \cd & \cd &     & \cd & \cd & \cd & \cd & \cd & \cd & \cd &     &     &     &  \\
        \bottomrule
        
    \end{tabular}
    }
    
    \caption{Metrics used for privacy protection technique comparison (\cd for metric used and \cb for metric used with breakdown).}
    \label{table:related_work_metrics}
\end{table*}

\subsection{Automated blocking list building}

Many privacy protection tools have been proposed (see \autoref{sec:protection_techniques}).
Most of these tools use manually maintained blocking lists (see \autoref{table:protection_techniques}). 
Some proposals automatically build tracking blocking lists using specific keys in URL that correspond to user identifying data \cite{Metwalley2015Unsupervised}, machine learning on DOM structure \cite{Bau2013Promising}, Javascript \cite{Wu2015Tracker,Ikram2017Towards} or user browsing behavior \cite{Kalavri2016Like}. The same machine-learning-based approach was also applied to ad-blocking list building using network traffic features \cite{Gugelmann2015Automated}.

\subsection{Privacy protection techniques comparison}    

While some extensions are used in almost all studies (e.g \adb \cite{AdBlockPlus} and \gho \cite{Ghostery}), many of them are seldom employed. 
Similarly, some works compare between four and seven extensions \cite{Mayer2012Third,Balebako2012Measuring,Hill2014Comparative,Hill2015uBlock,Wills2016What,Merzdovnik2017Block,Traverso2017Benchmark} while another focuses on two \cite{Krishnamurthy2007Measuring}.
\autoref{table:related_work_techniques} describes how existing web privacy-related work used or analyzed existing privacy protection techniques.

Features used to compare privacy protection techniques are very diverse and thus complicate result comparison across work.
Some works compare privacy protection techniques in terms of HTTP request number \cite{Mayer2012Third,Hill2014Comparative,Hill2015uBlock,Merzdovnik2017Block}, cookie number \cite{Mayer2012Third,Balebako2012Measuring,Hill2014Comparative,Hill2015uBlock}, domain number \cite{Hill2014Comparative,Wills2016What,Traverso2017Benchmark}, private information diffusion \cite{Krishnamurthy2007Measuring}, and occurrence of behavioral advertising \cite{Balebako2012Measuring}. 
Some counter-measure proposals are also comparing their approaches to others regarding tracker blocking \cite{Malandrino2013PrivacyAwareness,Malandrino2013PrivacyLeakage,Kontaxis2015Tracking} or impact on browser performance \cite{Achara2016My}.
\autoref{table:related_work_metrics} lists metrics and features leveraged by existing web privacy-related studies.

Our work is close to \cite{Krishnamurthy2007Measuring,Mayer2012Third,Balebako2012Measuring,Hill2014Comparative,Hill2015uBlock,Wills2016What,Merzdovnik2017Block,Traverso2017Benchmark} 
but improves the state of the art along four axes: protection techniques, target websites, metrics, and reliability. 
{\it First}, we compare more protection techniques (15 and additional combination of blocking lists), than Krishnamurthy et al.~\cite{Krishnamurthy2007Measuring} (2), Mayer et al.~\cite{Mayer2012Third} (4), Balebako et al. \cite{Balebako2012Measuring} (4), Hill \cite{Hill2014Comparative,Hill2015uBlock} (4), Wills et al. \cite{Wills2016What} (6), Merzdovnik et al. \cite{Merzdovnik2017Block} (5) and Traverso et al. \cite{Traverso2017Benchmark} (7).
It is difficult to compare the extensions covered by our work with Krishnamurthy et al. \cite{Krishnamurthy2007Measuring} since the ecosystem was much simpler at the time (AdBlock Plus and NoScript were the only extensions available).
We did not use some extensions that were addressed in previous studies because they are either, not supported anymore (e.g. TACO which was owned by Abine \cite{Balebako2012Measuring,Leon2012Why}), or not available for Firefox (AdBlock \cite{Metwalley2015The,Achara2016My,Wills2016What}, Superblock Adblocker \cite{Achara2016My}, Adremover \cite{Achara2016My} and Adblock Pro \cite{Achara2016My}).
Furthermore, AdBlock (resp Firefox Tracking Protection \cite{Kontaxis2015Tracking}) uses the same blocking list as AdBlock Plus (resp. Disconnect).
By analyzing AdBlock Plus and Disconnect, we thus provide performance bounds on these two techniques.
Achara et al. \cite{Achara2016My} and Wills et al. \cite{Wills2016What} also evaluated the performance of an ad-blocking extension that we do not address in this study: AdGuard Adblocker.
{\it Second}, we us more websites than most existing work.
We use the Alexa Top 1000 while Mayer et al.~\cite{Mayer2012Third}, Hill \cite{Hill2014Comparative,Hill2015uBlock} and Traverso et al.~\cite{Traverso2017Benchmark} use the Alexa Top 500, 45, 84 and 100 URLs, and 100 italian URLs.
Balebako et al. discussed five browsing scenarios that use a small number of websites to assess the occurrence of behavioral advertising, we thus cannot compare the websites we use.
We crawled a number of websites similar to Wills et al. \cite{Wills2016What} but analyzed many more protection techniques.
We use a smaller number of crawled websites than Merzdovnik et al. \cite{Merzdovnik2017Block} (who uses Alexa Top 100000) because some of our metrics (such as number of cookies) require a stateful crawl and thus forbid parallelism.
{\it Third}, we use more metrics (see \autoref{table:related_work_metrics}) than any existing work.
We are thus able to provide a better description of techniques' performance.
We did not compute the host number metric used by Hill \cite{Hill2014Comparative,Hill2015uBlock} because it is very similar to the number of domains.
This metric actually reflects the internal network architecture of trackers, which is not relevant for privacy protection comparison.
{\it Finally}, none of these work performs several measurements on each website to remove measurement error, except Mayer et al.~\cite{Mayer2012Third} who perform three crawls for each URL but does not provide any justification for this number.
To the best our knowledge, we here propose the first measurement error analysis for privacy protection technique comparison (see \autoref{sec:measurement_error}).

\section{Methodology}
\label{sec:methodology}

This section presents data collection and used metrics for privacy protection and webpage quality.

\subsection{Data collection}
\label{sec:data_collection}

We use OpenWPM \cite{Englehardt2016Online}, an open-source framework written in Python that relies on Selenium for browser automation.
OpenWPM supports Firefox and provides some extensions.
With minor modifications, we add extensions presented in \autoref{table:protection_techniques} and \autoref{sec:protection_techniques}.
This study has been conducted with Firefox 45.

We crawl the highest-ranked websites by Alexa \cite{Alexa}.
Unless stated otherwise, we perform measurements in August 2017, using IP addresses located in Japan.
Typically, a crawl on the Alexa Top 1000 (\autoref{subsec:extensions}) takes about three days with commodity hardware.

\subsection{Privacy protection}
\label{sec:metrics}

In this section, we present our robust methodology to compare privacy protection techniques across several websites and using several metrics.
We also present privacy footprint \cite{Krishnamurthy2006Generating}.

\subsubsection{Browsing metrics}
\label{sec:browsing_metrics}

As explained in \autoref{sec:web_tracking}, privacy leakage occurs through communications with trackers, and local data storage (e.g. cookie) allows trackers to easily identify users across website.
We consider five simple browsing-related metrics that reflect the ability of protection techniques to hinder these two phenomena. 
We first focus on HTTP requests.
We separately count the requests that are made to the accessed domain (\emph{the number of first party requests}), and the requests that are made to other domains (\emph{the number of third party requests}).
To this end, we use the second level domain name obtained from the Public Suffix List \cite{PublicSuffixList}.
These metrics give a rough estimation on the performance of a particular extension. 
A protection technique is effective if the number of first party requests is not impacted, and the number of third party requests decreases. 

The third metric is \emph{the number of third party domains} accessed during a crawl.
This is complementary to the number of third party requests.
It provides a metric on the number of blocked entities.
There may be a few domains that generate a lot of requests, or a many domains that produce a few requests.
Efficient protection techniques reduce the number of third party domains.

This is however not sufficient since third party requests are not always used to track users, they can also provide content to users (e.g. media resources) or contact non-tracking third parties (e.g. website analytics).
The fourth metric is \emph{the number of the profile cookies} obtained from a stateful crawl.

Protection techniques with good performances diminish the number of cookies.

The last metric is \emph{the total amount of data transferred}.
It is especially important in low-bandwidth situations for example on mobile.
This metric is directly related to tracking and advertisement. 
Advertisement often generates considerable traffic during browsing. 
However, as we will see in \autoref{sec:measurement_error}, this metric shows high variability when crawling the same website. 
We thus did not use this metric in our analysis.

\begin{figure}[]
  \centering
  \includegraphics[width=0.9\columnwidth]{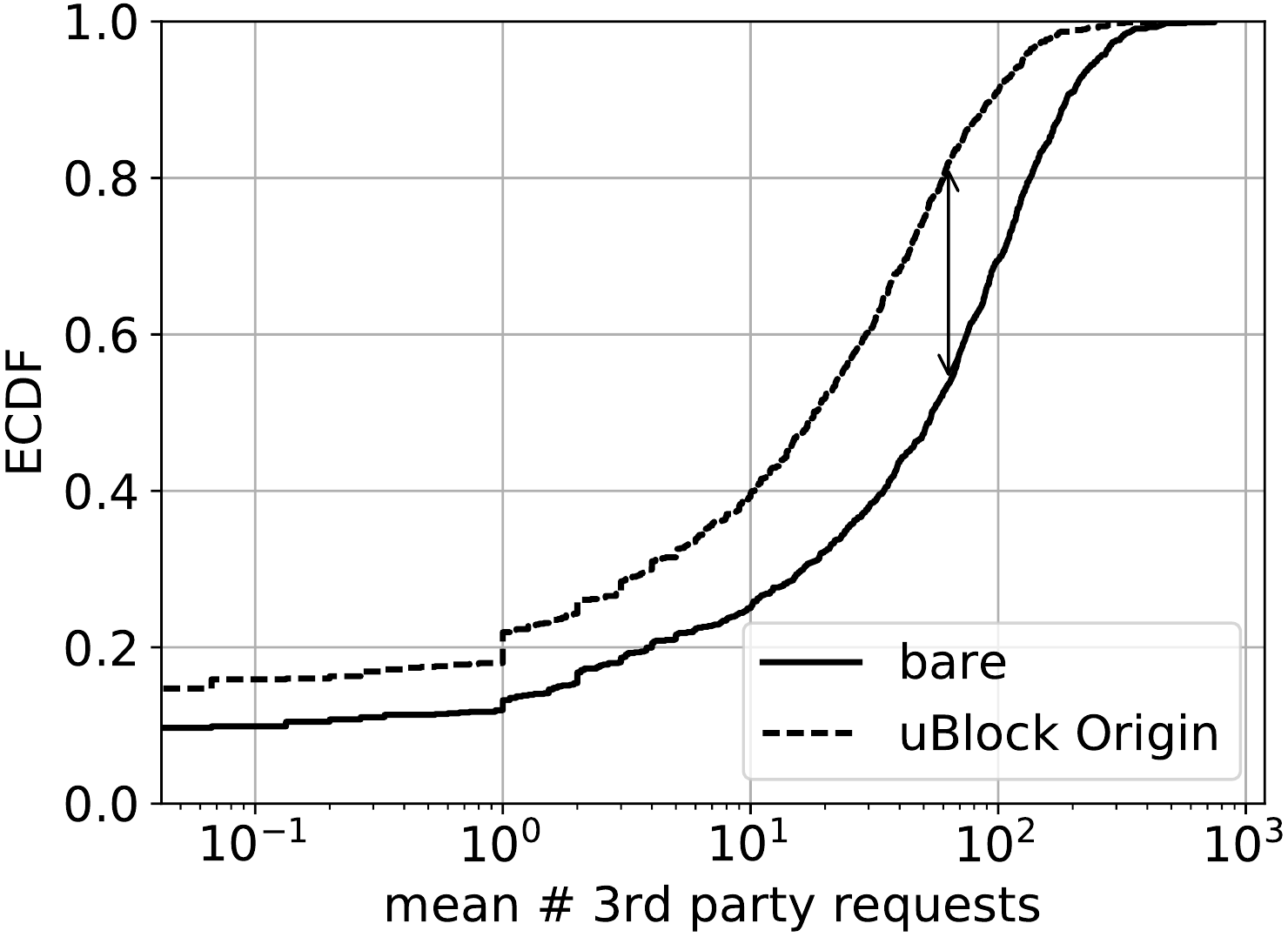}
  \caption{Example of ECDF of the mean number of third party requests for all websites in the Alexa Top 1000 with two configurations: default Firefox (bare) and Firefox with uBlock Origin. The arrow represents the KS statistic.}
  \label{fig:ecdf_example}
\end{figure}

\subsubsection{Kolmogorov-Smirnov test-based browsing metrics comparison}
\label{sec:ks_test}

While the metrics introduced in \autoref{sec:browsing_metrics} provide a good estimation of the privacy protection for a particular website, 
summarizing this aspect for a set of websites is non-trivial.
\autoref{fig:ecdf_example} presents two empirical cumulative distribution functions (ECDFs) built on the mean number of third party requests sent, for each website of the Alexa Top 1000 world.
We use the mean of each metrics for ten crawls to reduce measurement error (see \autoref{sec:measurement_error}).
We here use two Firefox configurations: one without any extension (bare) and one with uBlock Origin.
To determine whether one Firefox configuration exhibits the same performance as another, we use the Kolmogorov-Smirnov (or KS) test.
It is a non-parametric test that does not make any assumption on the underlying distributions which fit our context.
This test relies on the KS-statistic which is the maximum difference between two cumulative distributions.
The KS statistic is represented by the arrow on \autoref{fig:ecdf_example}. 
The null hypothesis is that both ECDFs belong to the same distribution.
In our case, this means that both configurations have the same performance.
We use a significance level $\alpha$ of 0.05.
If the p-value of the KS-test is smaller than $\alpha$, we consider the two ECDFs as distinct.
In other words, the two considered configurations have performances that are statistically different.

\subsubsection{Privacy footprint}
\label{sec:privacy_footprint}

In this work, we also use the privacy footprint proposed by Krishnamurthy et al. \cite{Krishnamurthy2006Generating}. 
The privacy footprint represents interactions between first parties and third parties (i.e. potential trackers) in a graph.
For each third party accessed by the user when visiting a first party, an edge is added between the nodes representing the considered first and third parties.
Privacy footprint thus captures both the potential information leaking from first parties, and the aggregating behavior of third parties.
Krishnamurthy et al. \cite{Krishnamurthy2006Generating} used the second-level domain name of the authoritative DNS server of third parties to group domains in the same entity on the same node in the graph.
This method is called \emph{ADNS}.
Krishnamurthy et al. \cite{Krishnamurthy2009Privacy} then noticed that unrelated third parties located on the same hosting service or Content Delivery Network (CDN) are grouped together by \emph{ADNS} because they share authoritative DNS servers.
They thus develop a new method called \emph{root}.
With \emph{root}, if a third party is located in a hosting service or a CDN, the second-level domain-name of the third party domain is the node identifier.
Otherwise, as with \emph{ADNS}, the second-level domain name of the authoritative DNS server of the considered third party is the node.

We use three metrics that are built on the graph.
(1) the \emph{number of third parties} reveals tracking breadth.
(2) the \emph{mean number of third parties per first party} corresponds to tracking intensity.
(3) the \emph{number of first parties associated with the top 10 third parties} estimates how much tracking is concentrated on the most prominent third parties.

\subsection{Webpage quality}
\label{sec:webpage_quality_methodology}

Privacy protection techniques prevent privacy information leakage by hindering of communications with trackers and blocking cookie creation, among other techniques.
This may however have negative side-effects on webpage quality.
Blocked third party domains may actually host images or JavaScript that are needed to render the webpage correctly.
We first analyze the impact of privacy protection on browsing data in an automated fashion.
We then perform a manual analysis to assess privacy protection repercussion on rendered webpage.

\subsubsection{HMTL-based metrics}
\label{sec:html_metrics}

Our goal is to detect layout differences or missing elements that may reduce webpage quality both in terms of rendering quality and functionality.
We crawl HTML data using OpenWPM.
We devise several metrics that analyze the browsed page.
The first metric is the \emph{HTML page size}.
We then derive two metrics from the crawled HTML data itself: \emph{number of images and scripts}.
The last two metrics are the \emph{total size of images and scripts}.
For all metrics, a reduction is associated with a webpage quality decrease.
We here reuse the metric comparison method presented in \autoref{sec:ks_test}.

\subsubsection{Manual analysis}
\label{sec:manual_analysis}

By analyzing HTML, we are able to determine how many elements and how much data is missing when privacy protection is used.
This, however, does not assess how well the webpage is rendered inside the browser.
We thus perform a manual analysis to determine webpage quality obtained with privacy protection techniques.
Webpage screenshots are captured using OpenWPM.
First, we want to determine whether privacy protection techniques impact webpage layout.
Our first question thus is: ``Please rate layout similarity between Bare (left) and Extension (right).''.
Good layout similarity, however, does not guarantee lack of missing elements.
The next step is to compare webpage elements when privacy protection is used and not used.
We want to avoid comparing elements of the same webpages that may change for different users or crawling time (e.g. news items).
We thus ask user to focus on webpage elements that are part of the user interface.
Our second question thus is: ``Please rate the proportion of elements (image, text frame, widgets, etc.) of the user interface (e.g.: login button, tabs, etc, but not news items or pictures) in Bare (left) that are also in Extension (right)''.
For both questions, users give a rating between 0 and 10, 0 being the worst and 10 the best.
Our manual analysis was performed by six users.

\section{Measurement parameters}

We address two measurement parameters: the impact of crawling parameters on measurement error, and the training of Privacy Badger.

\subsection{Impact of crawling parameters on measurement error}
\label{sec:measurement_error}

\begin{figure}[t]
  \centering
  \includegraphics[width=\columnwidth]{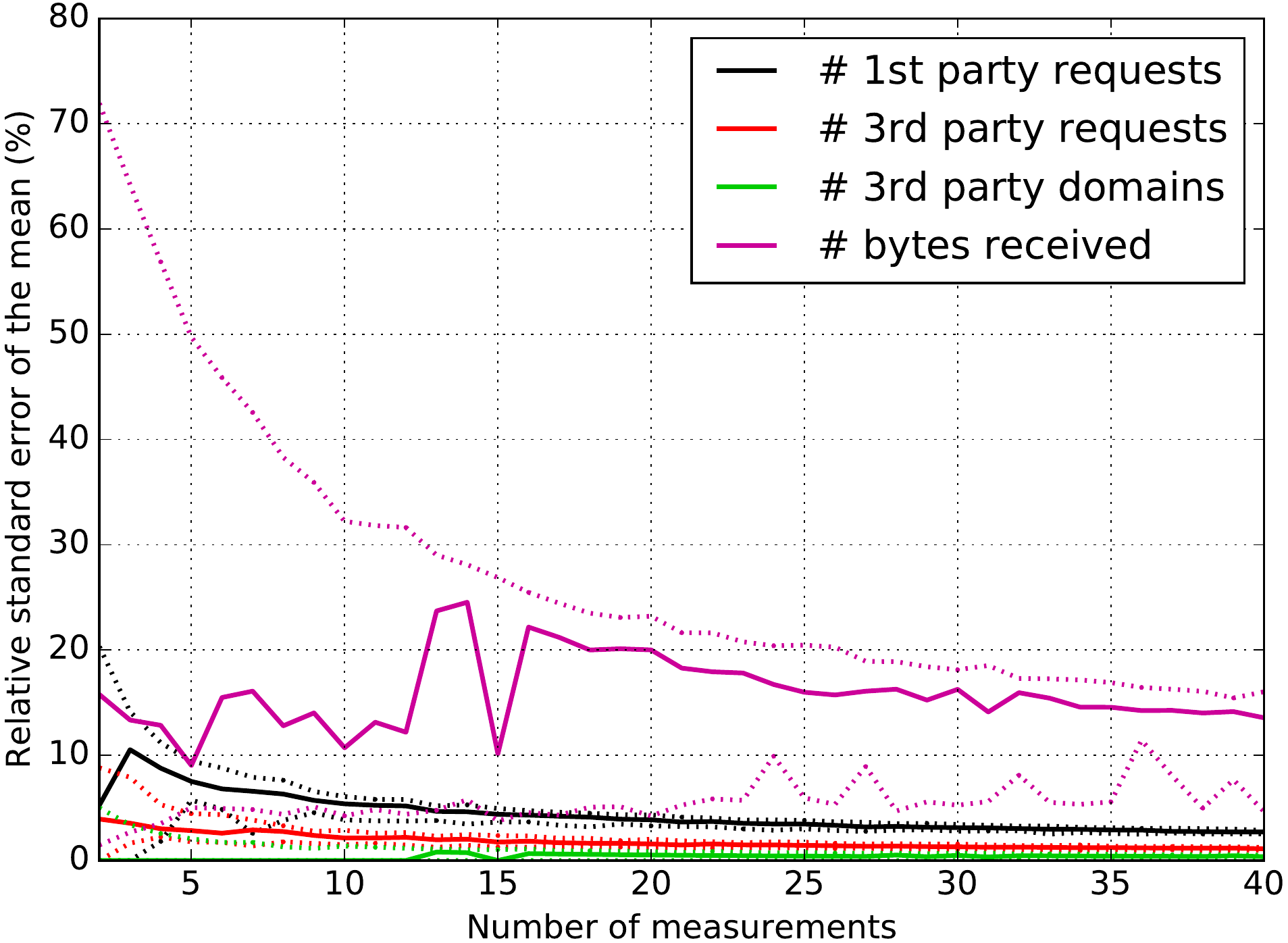}
  \caption{Relative standard error of the mean of browsing metrics in function of the number of measurements when accessing www.yahoo.jp. Solid lines represent the median, dashed lines correspond to 5 and 95-centiles.}
  \label{fig:measurement_error}
\end{figure}

During our preliminary experiment, we notice that measurement results are not stable.
Querying twice a website usually yields two different metric values.
To determine the appropriate number of measurements that generates a small error, we conducted a detailed study on the website showing the highest variability in our preliminary experiment: www.yahoo.jp.
These measurements were performed is November 2016.
Varying the number of measurements, we computed the relative standard error of the mean of four browsing metrics.

\begin{figure}[t]
  \centering
  \includegraphics[width=\columnwidth]{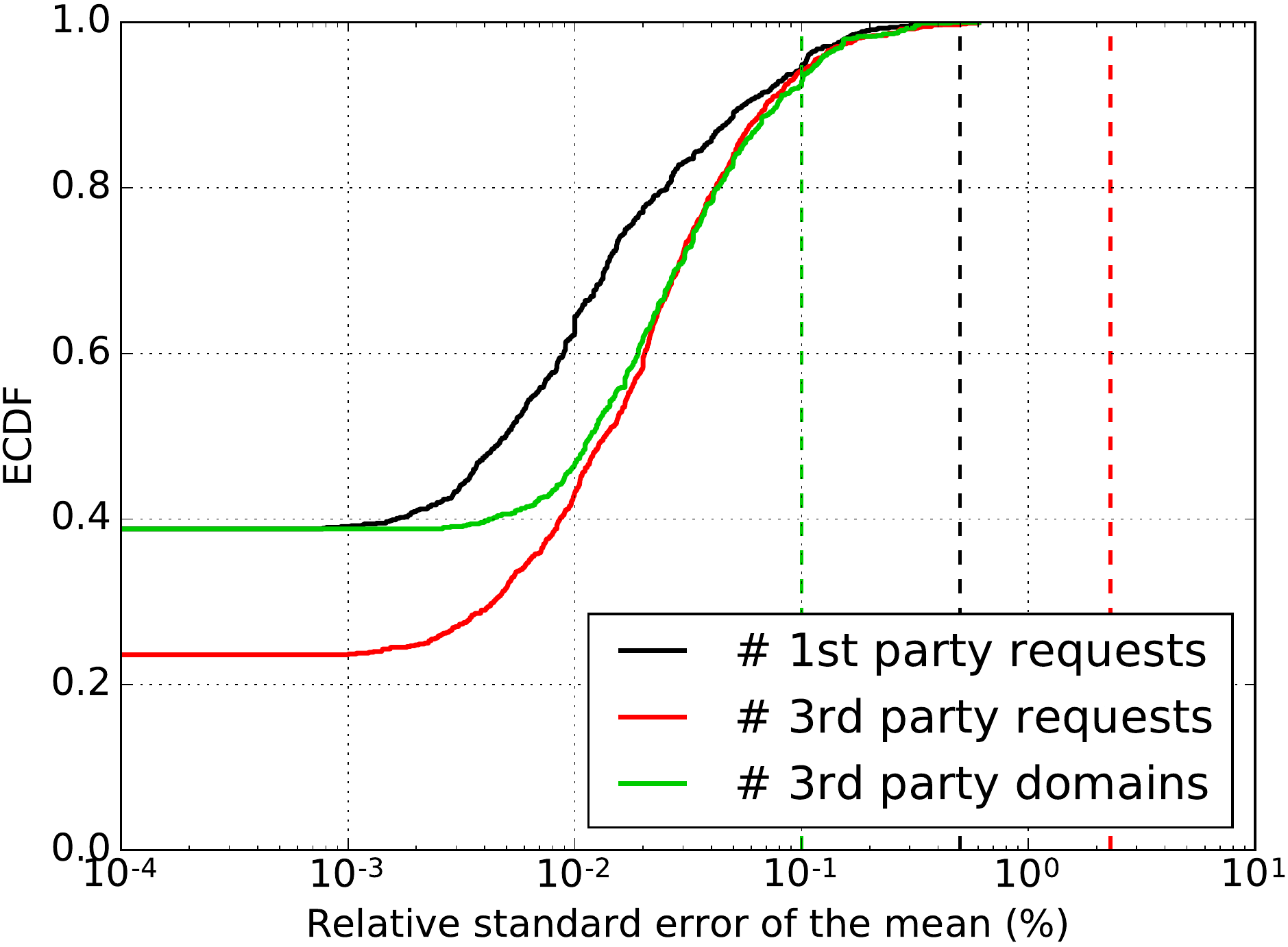}
  \caption{ECDF of the relative standard error of the mean of browsing metrics on the Alexa Top 1000 websites crawled ten times. Vertical dashed lines represent the relative standard error for www.yahoo.jp.}  
  \label{fig:measurement_error_top_1000}
\end{figure}

\autoref{fig:measurement_error} shows that the relative standard error decreases when the number of measurements increases. 
Performing ten crawls reduces the relative standard error on the number of first party requests, third party requests, and third party domains to less than 5\%.
The number of bytes received, however, still has a large error, we thus discard this metric.
One possible reasons for this measurement error is the webpage advertisement churn.
Guha et al. \cite{Guha2010Challenges} analyze ad churn during page reloads.
They show that the number of unique new ads increases quickly during the first ten reloads but only linearly thereafter.
This further justifies our choice to use ten measurements.
An existing comparison of privacy protection techniques \cite{Mayer2012Third} uses three crawls for each website and extension.
They however do not motivate this choice nor provided measurement error analysis.

\autoref{fig:measurement_error_top_1000} is the ECDF of the relative standard error of the mean of the number of first and third party requests, and of the number of third party domains for all websites in the Alexa Top 1000 crawled 10 times.
Most websites have a relative standard error smaller than that of www.yahoo.jp (here in dashed lines).
Overall, 99\% of websites have a relative standard error smaller than 1\% for all observed features.

\subsection{Privacy Badger training}
\label{sec:pb_training}

Privacy Badger employs heuristics to determine if a domain is performing tracking (see \autoref{sec:protection_extensions}).
Privacy Badger measures the number of times that a domain reads a cookie as a third party.
When this number is greater than three, the considered domain is blocked.
On top of this, cookies are also whitelisted using heuristics.
This behavior causes a freshly installed Privacy Badger not to block any domain.
As the user browses websites, more and more domains are blocked. 
To fairly evaluate Privacy Badger, we analyze the impact of \pb training.
In other words, we intend to quantify how many websites need to be accessed to ensure that \pb is completely trained.

\begin{figure}[t!]
  \centering
  \includegraphics[width=\columnwidth]{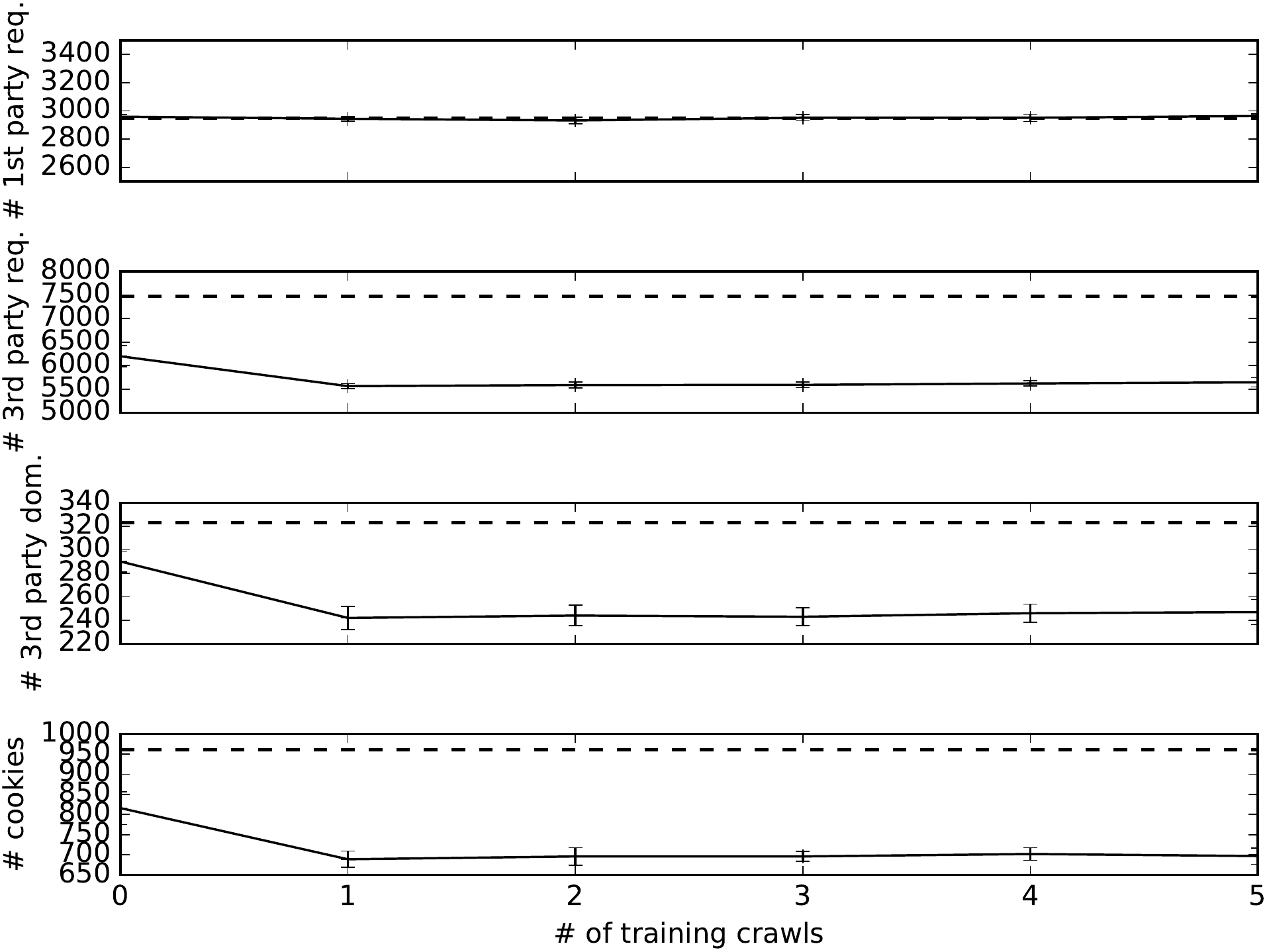}
  \vspace{-8mm}
  \caption{Performances of \pb depending on the number of crawl performed for its training. The horizontal dashed line represents a default Firefox as reference.}
  \label{fig:privacy_badger_training}
\end{figure}     

If we were to determine the number of websites regular users need to browse to train Privacy Badger, we should use a browsing model such as AOL search logs \cite{Pass2006Picture} as Roesner et al. \cite{Roesner2012Detecting} do.
Our use-case here is however to ensure that \pb is trained for a specific set of websites.
We thus measure Privacy Badger's performance on the Alexa Top 100 websites after training.
These crawls were performed is November 2016.
This training consists of several repeated crawls targeting the same set of websites and performed without reinitializing the user profile, i.e. \pb continues its training.
We use zero to five successive training crawls and then perform a single measurement crawl.

The results of this experiment are shown in \autoref{fig:privacy_badger_training}.
The dashed line corresponds to a reference: crawling results of a bare browser.
KS test groups the six measurements with Privacy Badger together, but separates them from the bare browser.
Without training, \pb is already able to provide partial protection.
After a single training pass, the number of third party requests is reduced by 10.3\%.
The results however do not improve when the number of training crawl increases.
We also ran a single training crawl on the Alexa Top 500 websites and observed similar performance as with a single training crawl.
The training is thus complete with less than one crawl (here one hundred websites).
MyTrackingChoices uses the same principle but without Privacy Badger's cookie data-based white-listing heuristics.
One can thus expect MyTrackingChoices to train faster.
Moreover, MyTrackingChoices is provided with a small bootstrapping tracker list which should further reduce training time.
For the remainder of this paper, we train \pb using a single crawl on the considered Alexa Top websites.
Out of the four existing work that uses \pb \cite{Ikram2017Towards,Hill2014Comparative,Merzdovnik2017Block,Traverso2017Benchmark}, only two actually peform some training \cite{Hill2014Comparative,Merzdovnik2017Block}.
We here provide the first estimation of how much website needs to be visited to complete Privacy Badger's training.

\section{Results}
\label{sec:results}

\begin{figure*}[ht!]
	\centering
	\subfloat[\# first party requests]{
		\centering
		\includegraphics[width=0.5\textwidth]{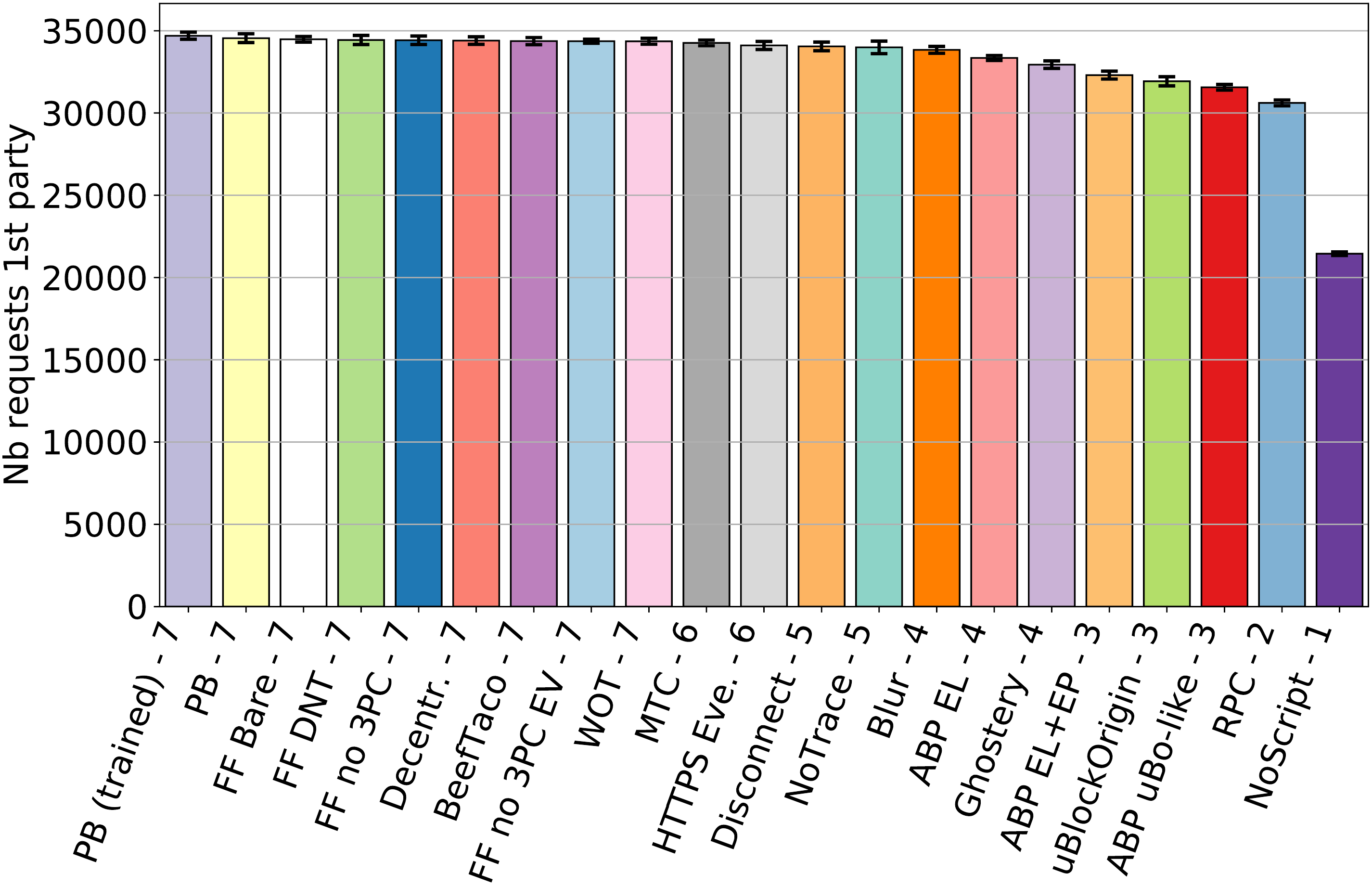}
		\label{fig:comparison_first_requests}
	}
	\subfloat[\# third party requests]{
		\includegraphics[width=0.5\textwidth]{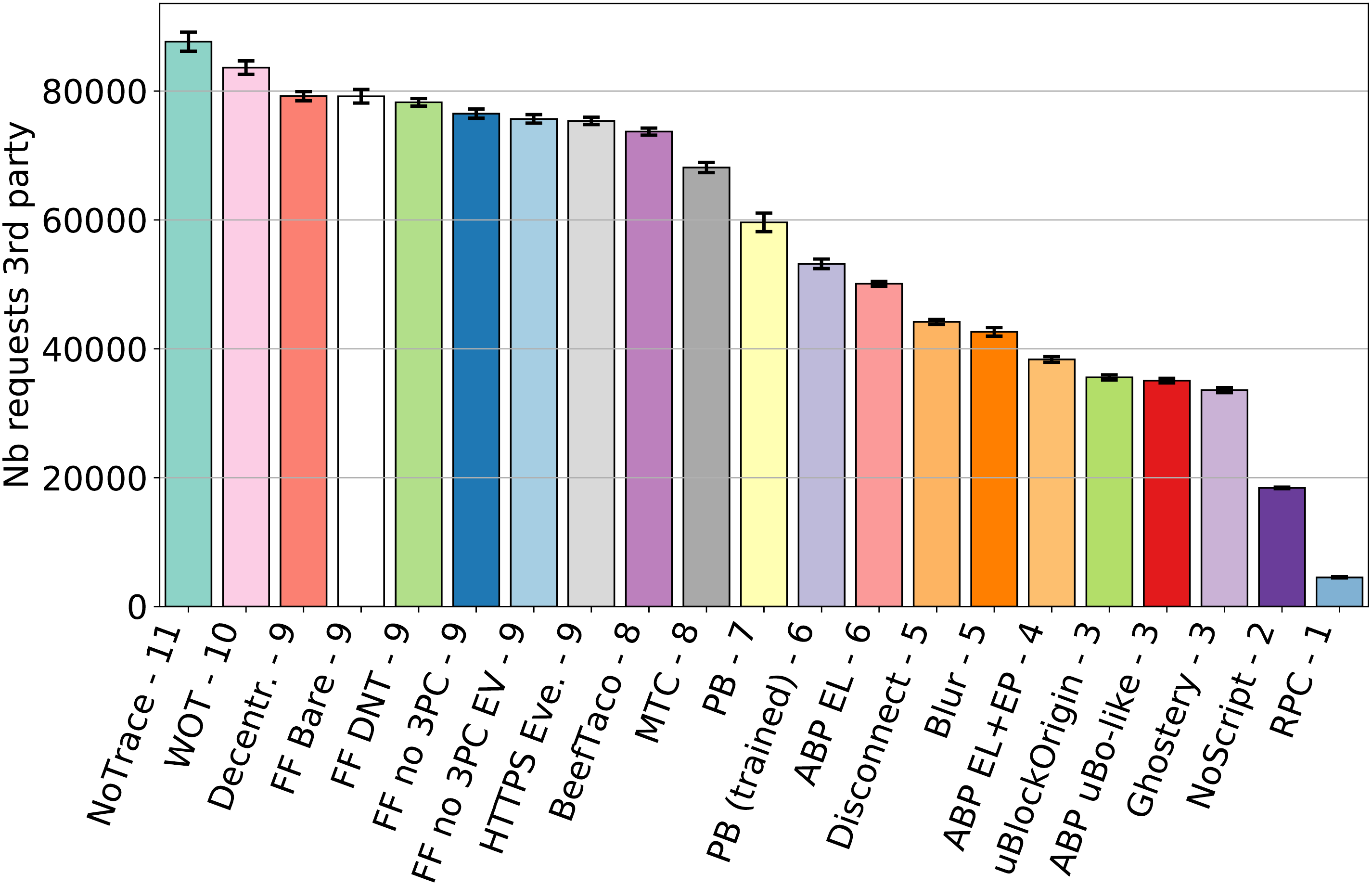}
		\label{fig:comparison_third_requests}
	}
	\\
	\subfloat[\# third party domains]{
		\includegraphics[width=0.5\textwidth]{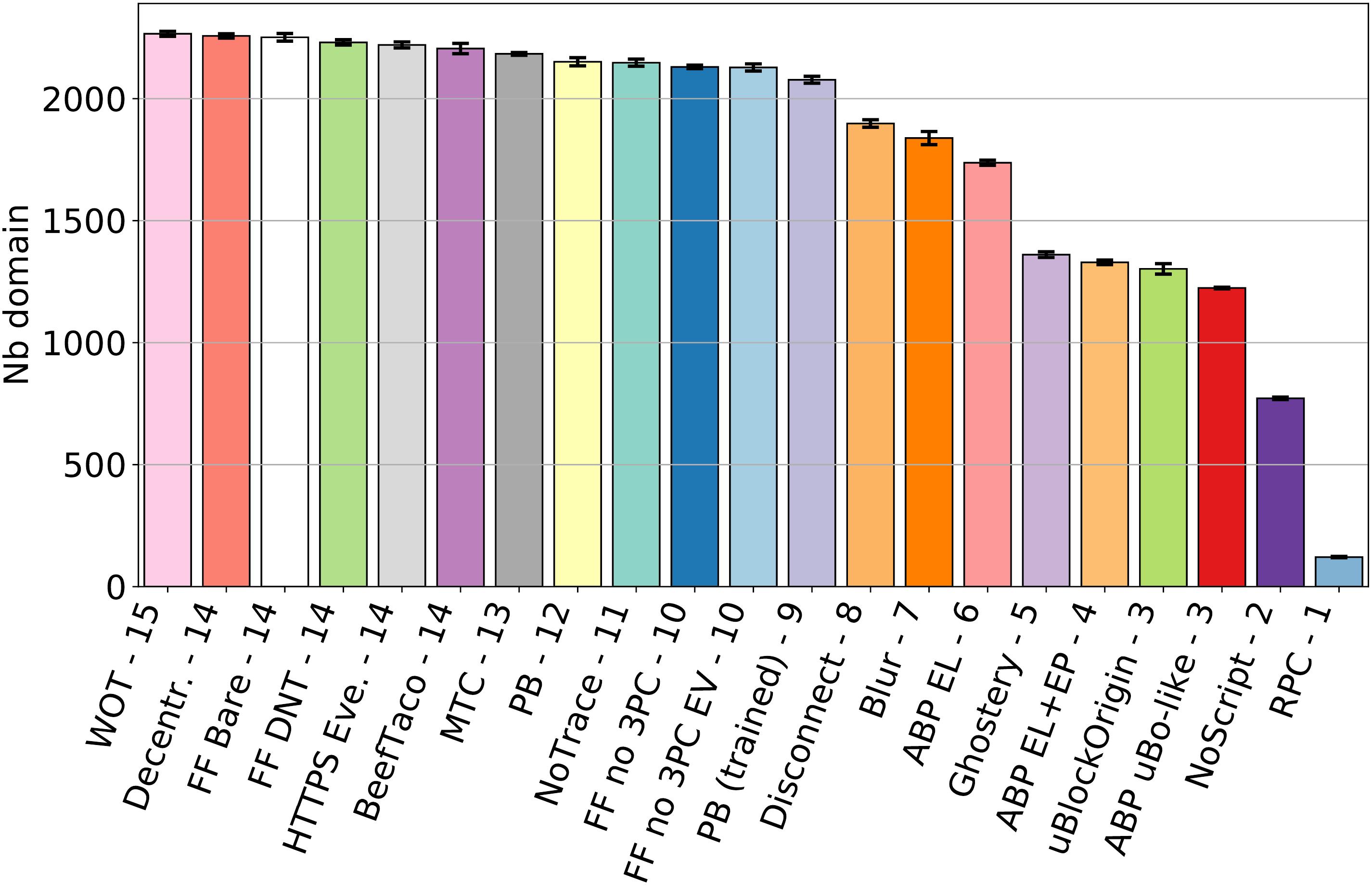}
		\label{fig:comparison_third_domains}
	}
	\subfloat[\# cookies]{
		\includegraphics[width=0.5\textwidth]{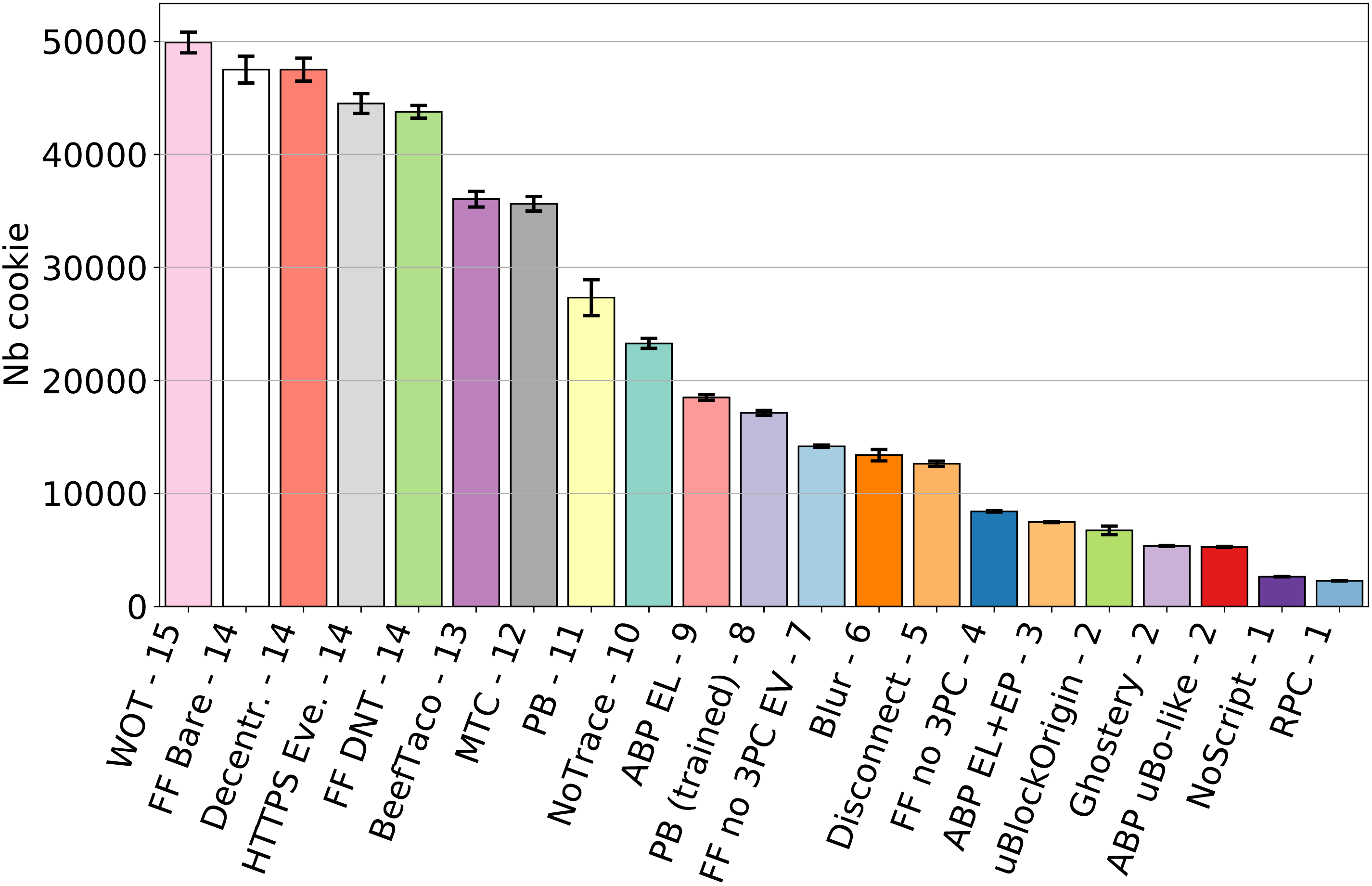}
		\label{fig:comparison_cookies}
	}
	\caption{Performance comparison of protection techniques regarding the mean \# first party request, mean \# third party requests, mean \# third party domains and \# cookies. Each metric value corresponds to the total obtained by crawling the Alexa Top 1000 websites. The mean and standard deviation are computed on ten crawls. The number on top of technique name is the KS-based rank. Bare: Firefox alone, MTC: MyTrackingChoices, PB: Privacy Badger, FF no 3PC: Firefox no 3rd party cookie, FF no 3PC EV: Firefox no 3rd party cookie blocking except from previously visited domains, ABP: AdBlockPlus, EL: EasyList, EP: EasyPrivacy, RPC: \rpc. AdBlock Plus uBo-like uses all uBlock Origin's lists except uBlock Origin's specific ones; it thus loads EasyList, EasyPrivacy, Peter Lowe's Ad Server list and Malware domains. }
	\label{fig:comparison_simple_metrics}
\end{figure*}

\subsection{Privacy protection}
\label{subsec:extensions}

We compare protection techniques, first in terms of performance and then, regarding their blocking overlap.

\begin{figure}[ht!]
  \centering
  \includegraphics[width=\columnwidth]{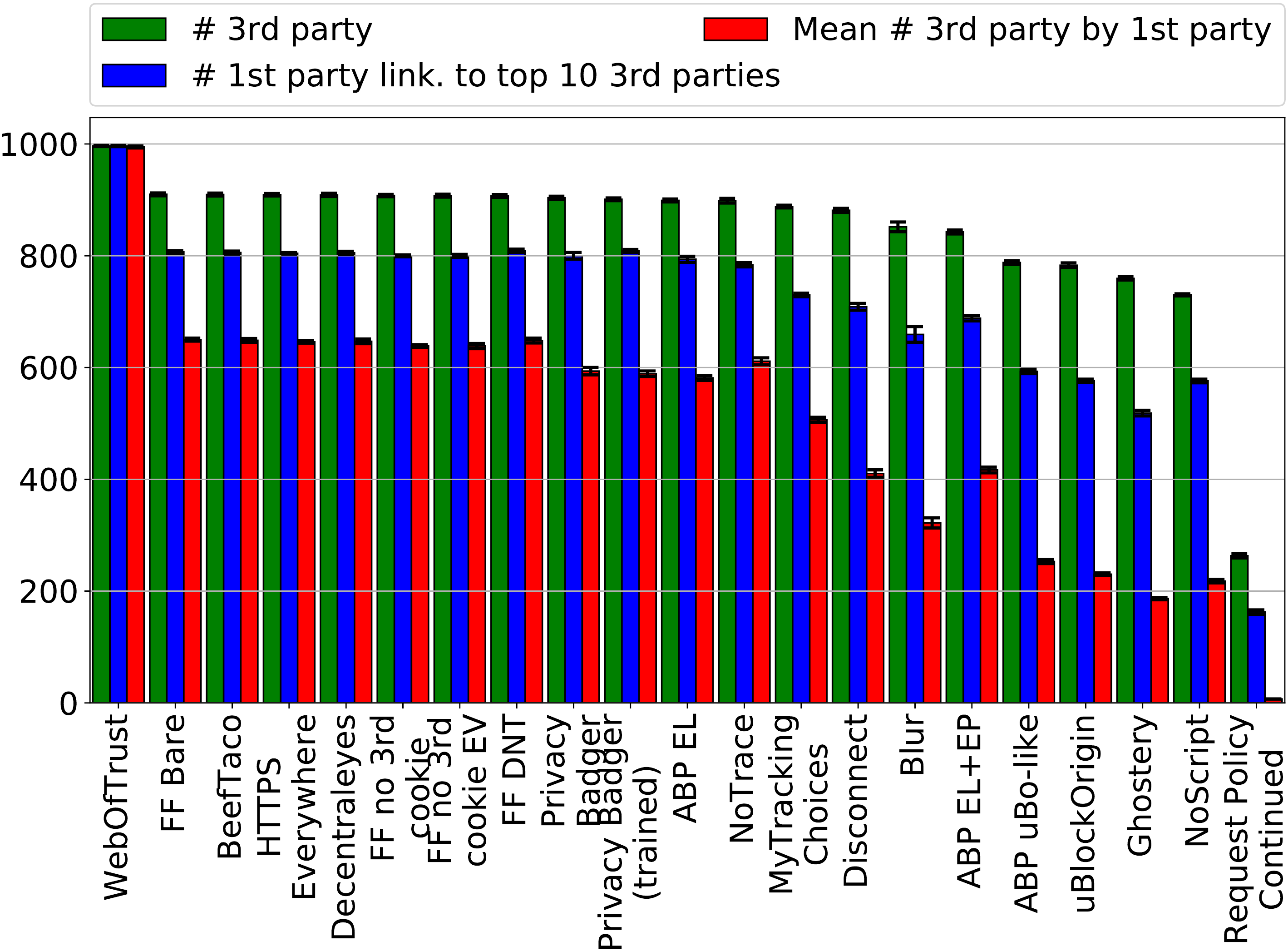}
  \caption{Privacy footprint \cite{Krishnamurthy2006Generating,Krishnamurthy2009Privacy} for the Alexa Top 1000 world. The mean and standard deviation are computed on the metric value of the ten crawl. EV: Firefox no 3rd party cookie blocking except from previously visited domains, ABP: AdBlockPlus, EL: EasyList, EP: EasyPrivacy.}
  \label{fig:comparison_privacy_footprint}
\end{figure}

\begin{figure*}[ht!]
  \centering
  \subfloat[\# third party requests]{
    \includegraphics[width=0.5\textwidth]{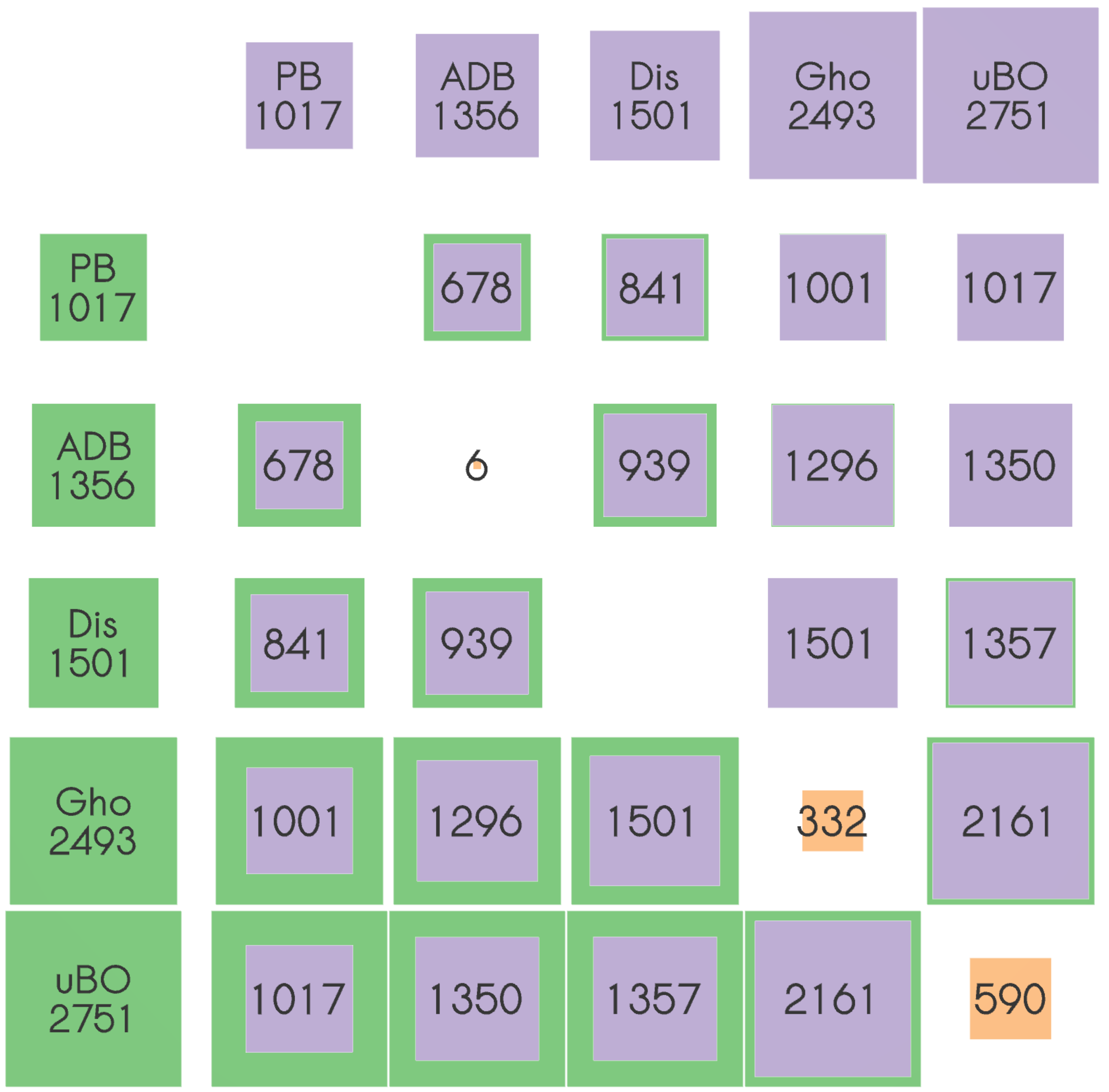}
    \label{fig:chord_3rd_party_requests_tulip}
  }
  \subfloat[\# third party domains]{
    \includegraphics[width=0.5\textwidth]{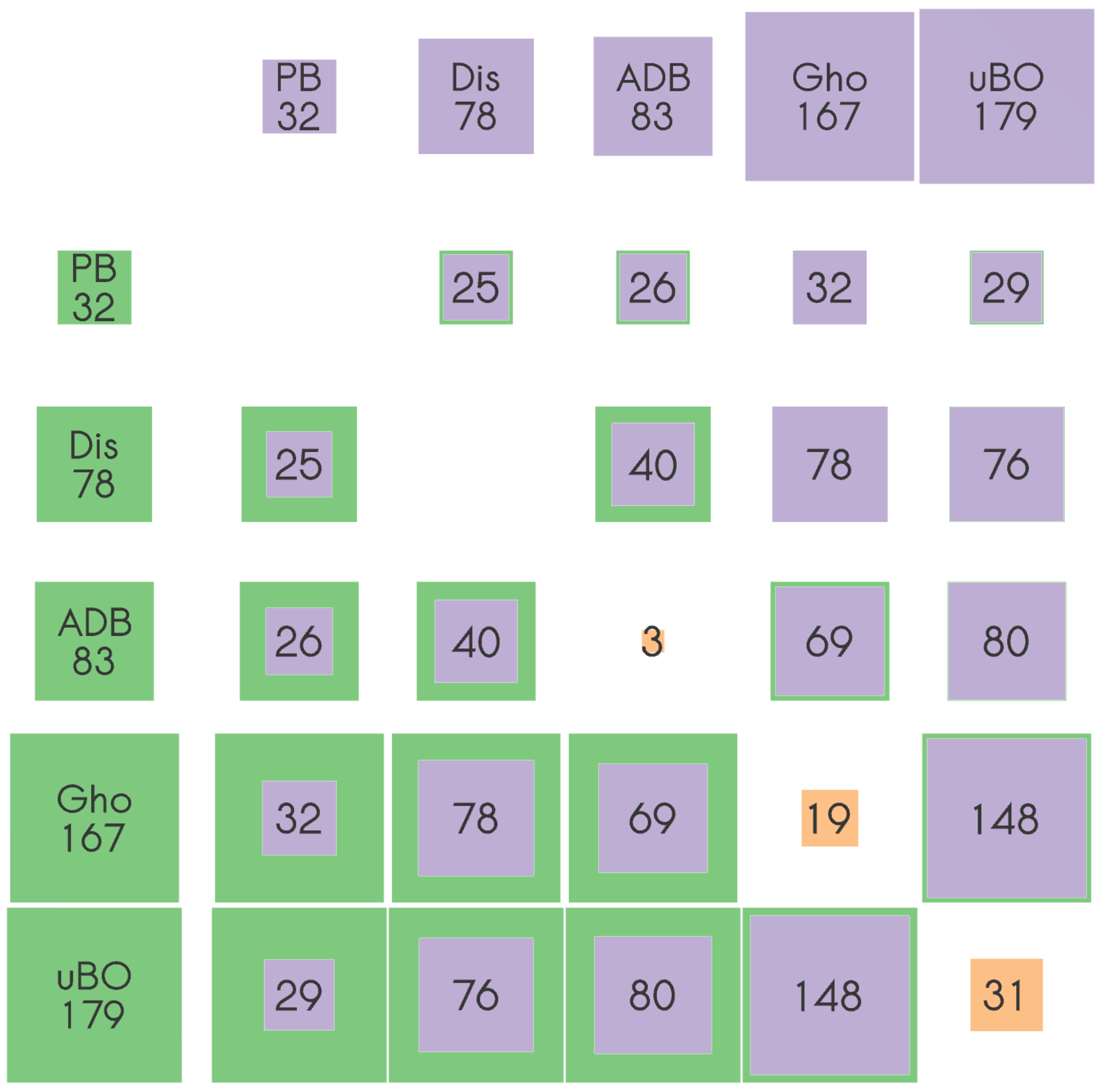}
    \label{fig:chord_3rd_party_domains_tulip}
  }
  \caption{Overlap of third party requests and domains blocked by five extensions (PB: Privacy Badger, Dis: Disconnect, ADB: AdBlock Plus, Gho: Ghostery, uBO: uBlock Origin).
    Square surfaces represent the metric value for the extension on the row.
  }
  \label{fig:extension_overlap}
\end{figure*}

\begin{figure*}[ht!]
	\centering
	\subfloat[\# images]{
		\includegraphics[width=0.5\textwidth]{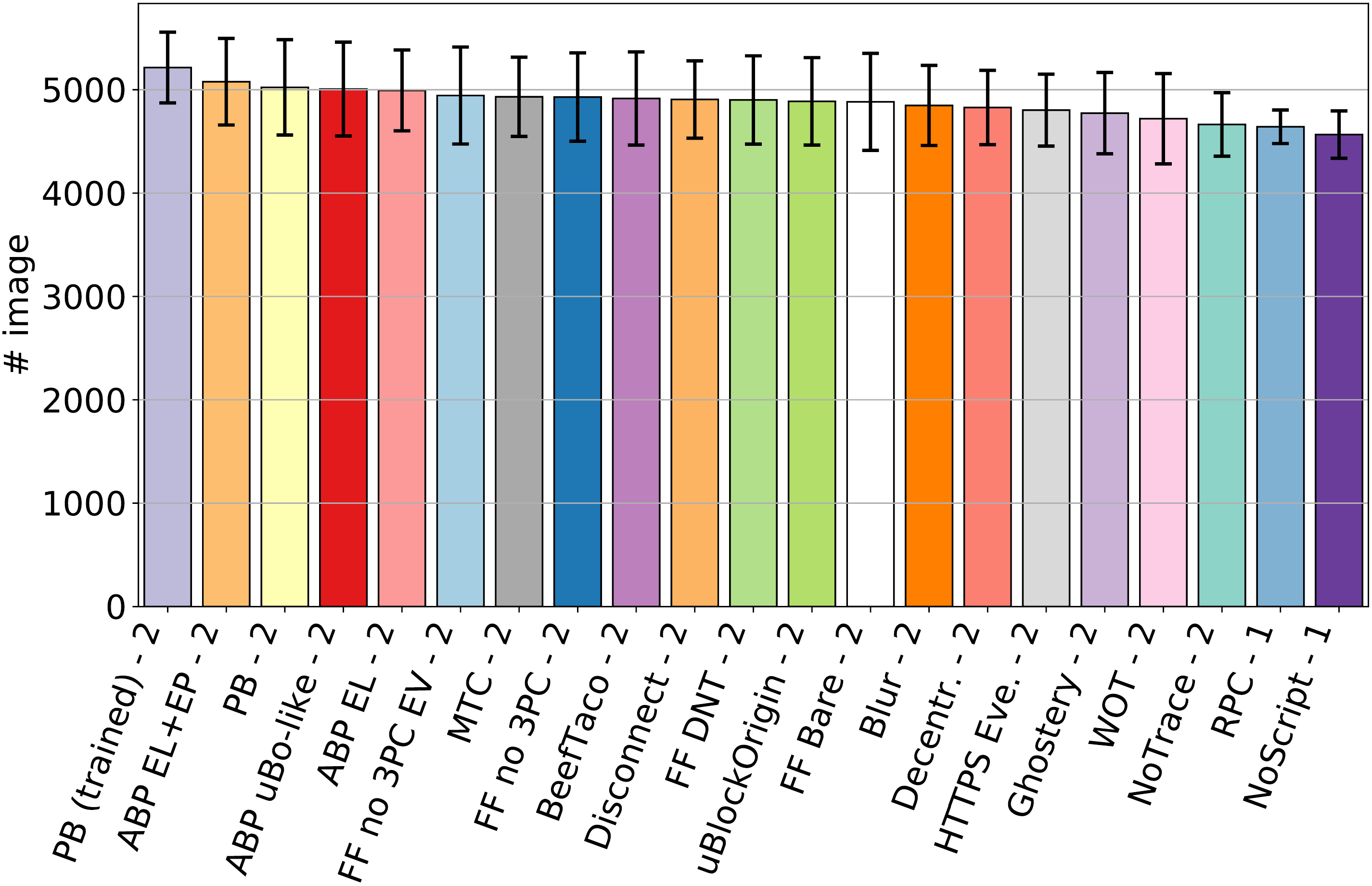}
		\label{fig:comparison_image}
	}
	\subfloat[Image size]{
		\includegraphics[width=0.5\textwidth]{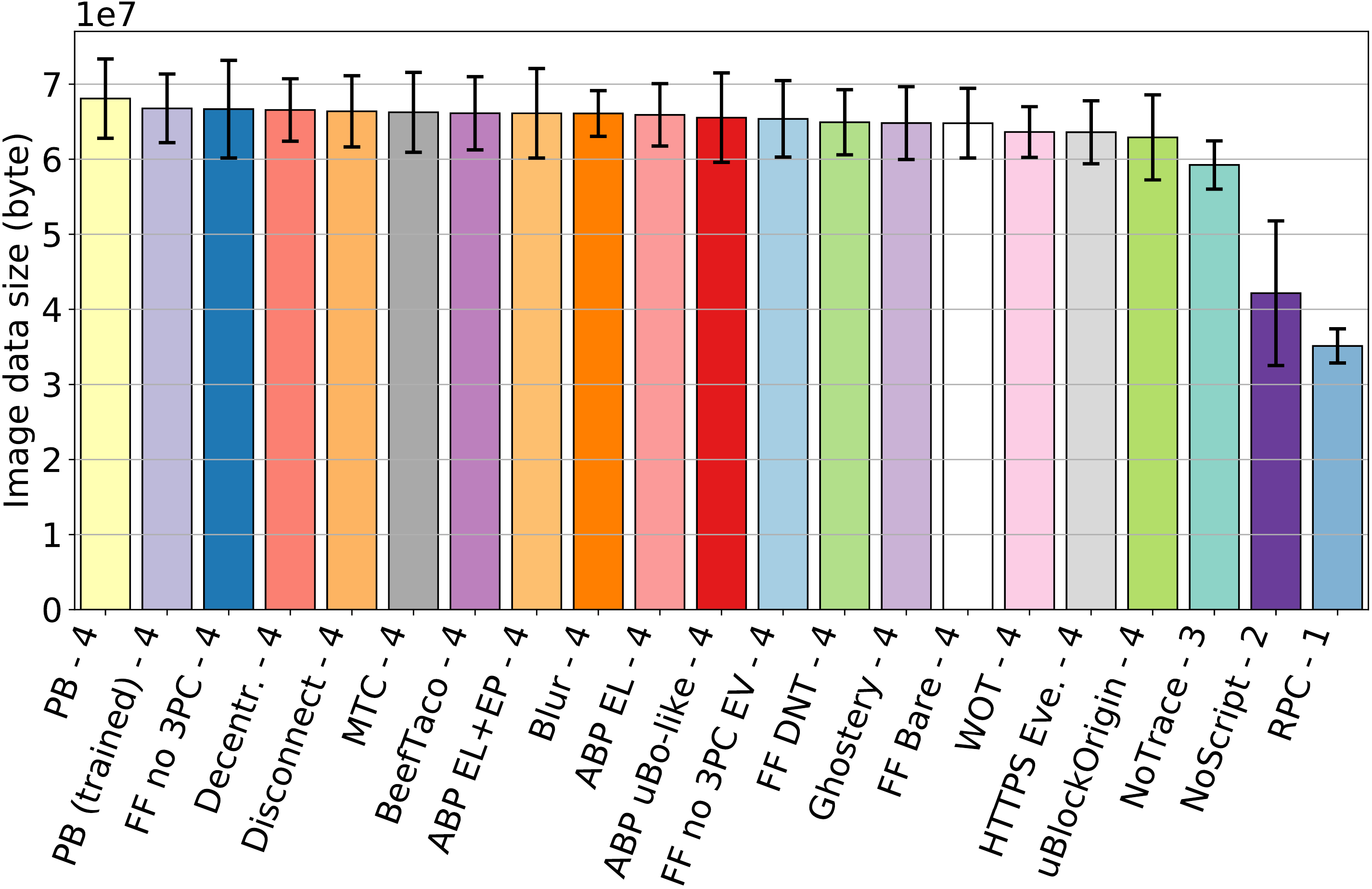}
		\label{fig:comparison_image_size}
	}
	\\  
	\subfloat[\# scripts]{
		\includegraphics[width=0.5\textwidth]{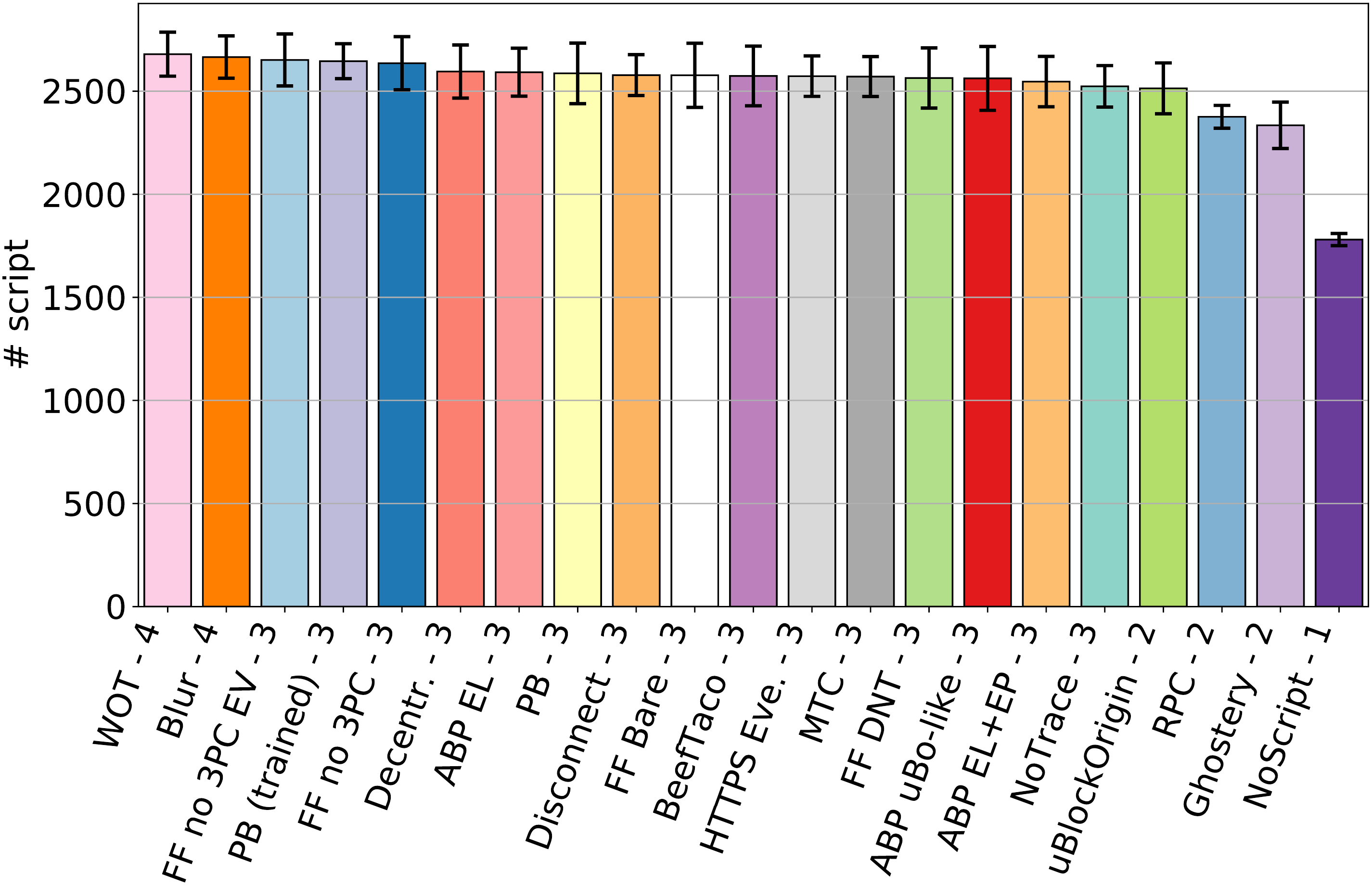}
		\label{fig:comparison_script}
	}
	\subfloat[Script size]{
		\includegraphics[width=0.5\textwidth]{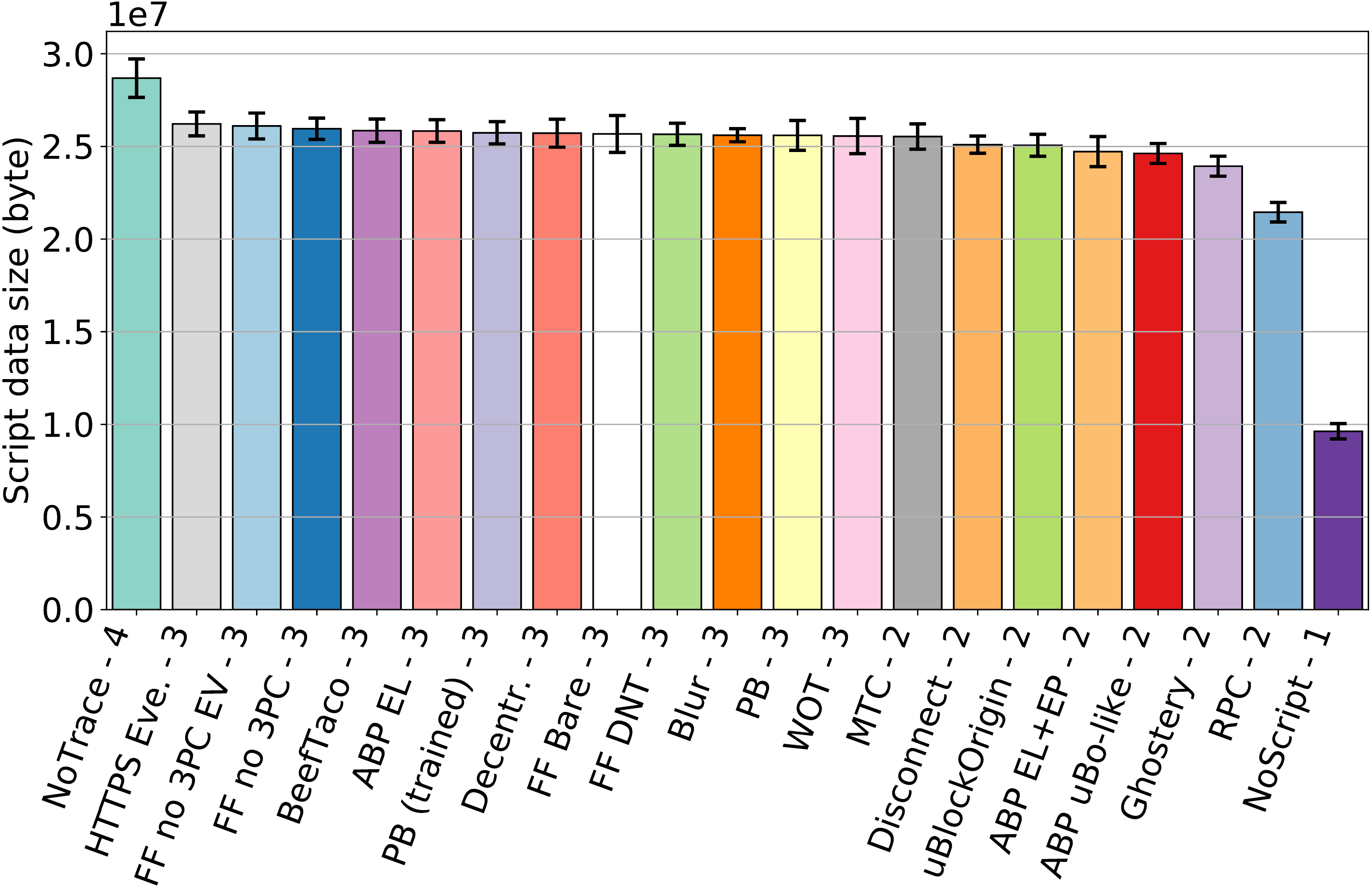}
		\label{fig:comparison_script_size}
	}
	\caption{Comparison of protection techniques' website quality regarding: number of image and scripts and size of images and scripts. We crawled the Alexa Top 100 websites. The mean and standard deviation are computed on ten crawls. The number on top of technique name is the KS-based rank. Bare: Firefox alone, FF: Firefox feature, MTC: MyTrackingChoices, PB (trained): Privacy Badger (trained), FF no 3rd coo. EV: Firefox no 3rd party cookie blocking except from previously visited domains, ABP: AdBlockPlus, EL: EasyList, EP: EasyPrivacy, RPC: Request Policy Continued. AdBlock Plus uBo-like uses all uBlock Origin's lists except uBlock Origin's specific ones; it thus loads EasyList, EasyPrivacy, Peter Lowe's Ad Server list and Malware domains.}
	\label{fig:comparison_website_quality}
\end{figure*}

\subsubsection{Browsing metrics}
\label{sec:comparison_simple_metrics}

We perform a general comparison of privacy protection techniques. 
We crawled the Alexa Top 1000 World using 21 different browser configurations.
One configuration is the default setup of the Firefox browser (Bare).
One configuration is Firefox setup with the DoNotTrack HTTP header.
Two configurations use Firefox's ability to block cookies: third party ones or third party cookies except from previously visited websites.
We do not use complete cookie blocking because we hypothesize that users want to keep using cookie for some domains.
Three configurations set up AdBlock Plus with various lists: EasyList, EasyList and EasyPrivacy and a filter set similar to uBlock Origin (EasyList, EasyPrivacy, Peter Lowe's Ad Server list and Malware domains).
One configuration uses NoTrace with the high preset (the default configuration has no effect on our metrics).
The last 14 configurations are extensions with their default setting.
We do not use the Firefox Tracking Protection \cite{Kontaxis2015Tracking} because OpenWPM does not support it yet.
However, this technique uses the Disconnect blocking list.
This means that by analyzing Disconnect, we provide a performance bound on Firefox Tracking Protection's performances.
The results are shown in \autoref{fig:comparison_simple_metrics}.
The number on top of technique name is the KS-based rank (see \autoref{sec:ks_test}).

All techniques have a limited impact on first party requests.
The most aggressive technique is NoScript \cite{NoScript}.
This is due to the fact that Javascript is pervasive on Internet.
Other techniques exhibit a limited impact on th enumber of blocked first party requests.

Except for the number of cookies, the two most effective extensions are \rpc and NoScript which reduce the number of third party HTTP requests by 96\% and 87\%.
However, as shown in \autoref{sec:webpage_quality} and other studies \cite{Krishnamurthy2007Measuring,Malandrino2013PrivacyLeakage}, they strongly impact webpage quality.

Among other techniques, \gho \cite{Ghostery} and \ubo \cite{uBlockOrigin} provide the best performances.
They reduce the number of third party HTTP requests by 58\% and 55\%.
Ghostery however needs to be configured which is difficult for users \cite{Leon2012Why}.
AdBlock Plus \cite{AdBlockPlus} uBO-like has similar performances with uBlock Origin and Ghostery.
Our results show that users can manually setup AdBlock to obtain good results, unfortunately this is difficult \cite{Leon2012Why}.
Furthermore, AdBlock Plus used with EasyList is much more CPU and memory-hungry than uBlock Origin \cite{Achara2016My}.

Another group of techniques provides average performances: Disconnect (and Firefox Tracking Protection), AdBlock Plus (with EasyList and EasyList + EasyPrivacy), Privacy Badger, Blur, BeefTaco and NoTrace.
Their results are mostly consistent across third party requests and domains, and cookies.
Privacy Badger \cite{PrivacyBadger}, \blur \cite{Blur}, Disconnect \cite{Disconnect} and AdBlock Plus with the EasyList block a relatively large number of third party HTTP requests (between 33\% and 48\%), but a fairly small number of third party domains (between 8\% and 33\%).
We hypothesize that these techniques block the main trackers but fail to impact the tracking long-tail \cite{Englehardt2016Online}.
MyTrackingChoices performances are worse than untrained Privacy Badger.
We hypothesize that this is due to a GUI bug that forbids users from customizing website categories where blocking occurs.
MyTrackingChoices thus always block tracking on 13 websites categories out of 32.
BeefTaco has a noticeable impact on cookies but no impact on other metrics.
This is consistent with previously observed tracking long tail \cite{Englehardt2016Online}.
This phenomenom makes the opt-out cookie maintenance very difficult while the protection that they offer is limited to cookie blocking.
NoTrace \cite{NoTrace}  provides average protection despite being setup at its highest level.
NoTrace increases users' online privacy awareness \cite{Malandrino2013How}, but unfortunately, users cannot reliably setup the extensions \cite{Leon2012Why} which is required.

Remaining techniques provide poor protection.
Decentraleyes \cite{Decentraleyes} has almost no effect on the results.
We hypothesize that blocking library loading requests only marginally impacts HTTP traffic.
The DoNotTrack HTTP header \cite{DoNotTrack} also has barely no effect.
If trackers complied with DoNotTrack, the number of cookies would significantly drop.
We thus conclude that DoNotTrack is not respected by most trackers.
WebOfTrust and HTTPSEverywhere have no effect.

We note that WebOfTrust and NoTrace both yield significantly more third party requests than Firefox without any extensions.
WebOfTrust also contacts more third party domains and increases the number of cookies.
WebOfTrust annotate hyperlinks in webpages with trust rating.
We hypothesize that this rating is obtained from WebOfTrust servers and thus impact our metrics.
We do not have any explanation regarding NoTrace's behavior.

When we block third party cookies in Firefox, there is a reduction in the number of third party domains.
Blocking cookies also slightly reduces the number of third party requests.
As hypothesized in \cite{Englehardt2016Online}, this may be due to impeded cookie-syncing interactions.
However, contrary to Englehardt et al. \cite{Englehardt2016Online}, third party cookie blocking here has poor performances.
This may be due to the fact that, unlike \cite{Englehardt2016Online}, we perform a stateful crawl. 
In a previous study \cite{Englehardt2015Cookies} that uses stateful crawl and OpenWPM, third party cookie blocking also exhibits poor performances.
The metric used in this study is however extremely specific, it thus is very difficult to compare our results with theirs.
Furthermore, some third parties from the Alexa Top 1000 are first party at some point (e.g. Twitter, Facebook).
Their cookies are thus created as first parties.

\subsubsection{Privacy footprint}
\label{sec:comparison_privacy_footprint}

We then compared extensions using the privacy footprint (see \autoref{sec:privacy_footprint} and \cite{Krishnamurthy2006Generating}).
\autoref{fig:comparison_privacy_footprint} presents these results.
They are similar to browsing metrics' \autoref{fig:comparison_simple_metrics}.
Six extensions exhibit much better results than the rest: RequestPolicy Continued, NoScript, Ghostery, uBlock Origin and AdBlock Plus with uBlock Origin's list.
For RequestPolicy Continued, the mean number of third parties per first party is very small while the total number of third parties is much higher.
This means that some third parties are detected but that they are present on a limited set of first parties.
NoScript results are here close to Ghostery's.
The hierarchy between Ghostery, uBlock Origin and the customized AdBlock Plus is here very clear and follows this order.
We hypothesized that the uniqueness of Ghostery and uBlock Origin blocking lists (see \autoref{sec:comparison_overlap}) is the main factor of their superior performances.

\subsubsection{Overlap}
\label{sec:comparison_overlap}

Next, we measure the overlap between the five extensions among the most popular ones regarding the number of blocked third party requests and blocked third party domains using a matrix view.
We here data crawled on the Alexa Top 100 websites in November 2016.
We do not analyze NoScript \cite{NoScript} (resp. \rpc \cite{RequestPolicyContinued}) here because it indiscriminately blocks Javascript (resp. third party requests).
Privacy Badger has been trained on the Alexa Top 100 websites and each extension is used with its default blocking list.
AdBlock Plus uses the EasyList.
uBlock Origin loads EasyList, EasyPrivacy, Peter Lowe's Ad Server list, Malware domains, and some uBlock specific lists.
Ghostery is set up with all its filter categories, while Disconnect is employed with its own filters.

Let us consider two extensions.
Let $E_i$ be the extension on the row $i$, and $E_j$ the extension on the column $j$.
The resources blocked by $E_i$ (resp. $E_j$) are noted $B_i$ (resp. $B_j$).
On the left (resp. top), the green (resp. purple) square on row $i$ (resp. column $j$) represents the total number of resource blocked by $E_i$ (resp. $E_j$).
On each line $i$ of the matrix, the surface of the square represents the total number of resources blocked by $E_i$: $\vert B_i \vert$.
The inner purple square on row $i$ and column $j$ represents the intersection between $E_i$ and $E_j$ and its surface is $\vert B_i \cap B_j \vert$.
The diagonal orange square surface displays resources that are only blocked by $E_i$: $\vert B_i - \{B_k \forall k \not= i\} \vert$.

As an example, we see AdBlock Plus on the second line of Fig.~\autoref{fig:chord_3rd_party_requests_tulip}.
Square surfaces on this line represent the number of third party requests blocked by AdBlock Plus: 1356.
The first square on the considered line represents the intersection with Privacy Badger which is encoded by the surface of the inner purple square, here 678.
The green surface outside the purple square describes the resources blocked by AdBlock Plus but not by Privacy Badger.
The next square on this line is located on the diagonal, and represents the resources blocked by AdBlock Plus only.
Remaining squares on this line depicts intersections with Disconnect, Ghostery and uBlock Origin in the same fashion as the intersection with Privacy badger is represented (see above).

The overlap of the third party requests (Fig.~\autoref{fig:chord_3rd_party_requests_tulip}) shows that there is a big overlap across all extensions.
All third party requests blocked by Privacy Badger are also blocked by uBlock Origin.
Similarly, Ghostery almost completely covers Privacy Badger.
Despite using two different blocking techniques, lists and heuristics, these extensions exhibit a significant overlap. This proves that the Privacy Badger's heuristics are reliable.

Results on the third party domains on Fig.~\autoref{fig:chord_3rd_party_domains_tulip} are similar to Figure~\autoref{fig:chord_3rd_party_requests_tulip}.
The major difference is that Privacy Badger's impact is smaller.
This is consistent with the heuristics behavior (see \autoref{sec:pb_training}) that only block third party domains that are seen across several websites.
These heuristics thus have trouble blocking less prominent trackers \cite{Englehardt2016Online}.

Both figures illustrate that Ghostery and uBlock Origin block many specific resources that other extensions do not block.
We speculate that this is due the fact that their blocking list settings are unique.
This may explain why they exhibit the best performance (see \autoref{sec:comparison_simple_metrics}).

Some resources are only blocked by AdBlock Plus.
It however uses EasyList which is also employed by uBlock Origin.
The ten crawls made with AdBlock Plus may have reached some third parties that were not contacted during the ten measurements made with uBlock Origin.
We thus hypothesize that this observation is another side-effect of the ad churn observed by Guha et al. \cite{Guha2010Challenges}.

\subsubsection*{Summary}

The most popular extensions show a wide overlap.
Ghostery and uBlock Origin block specific resources that are not affected by other extensions.
In terms of overall privacy protection, RequestPolicyContinued and NoScript show the best performances.
Ghostery and uBlock Origin protect users slightly less.
Remaining techniques provide average to low protection.
The DoNotTrack HTTP header provides almost no protection.

\subsection{Webpage quality}
\label{sec:webpage_quality}

We finally analyze how privacy protection techniques impact website quality.
We use an HTML-based automated approach and perform a manual analysis.

\subsubsection{HTML metrics}
\label{sec:comparison_html_metrics}

We crawl the Alexa Top 100 world on the 21 different browser configurations of \autoref{sec:comparison_simple_metrics}.
We use the metrics presented in \autoref{sec:html_metrics} and the ranking method exposed in \autoref{sec:ks_test}.
\autoref{fig:comparison_website_quality} presents the results.
We discard the HTML size metric because there is no significant change across protection techniques.
NoScript has a noticeable impact on all metrics.
This is especially visible for the script number, and image and script size.
RequestPolicy Continued's impact is smaller than NoScript's except for image size.
We hypothesize that \rpc blocks third party image hosting services, and thus impacts many websites.
NoScript greatly reduces the total size of scripts which is consistent with its goal: blocking JavaScript on webpages.
Other privacy protection techniques appear to have a limited impact of website quality.
Compared to other techniques, Ghostery exhibits a smaller impact on images and a higher reduction of the number of script and their total size.
This is consistent with its tracking protection goal.
This extension actually does not intend to block media such as advertisements.

\begin{figure}[t!]
  \centering
  \includegraphics[width=\columnwidth]{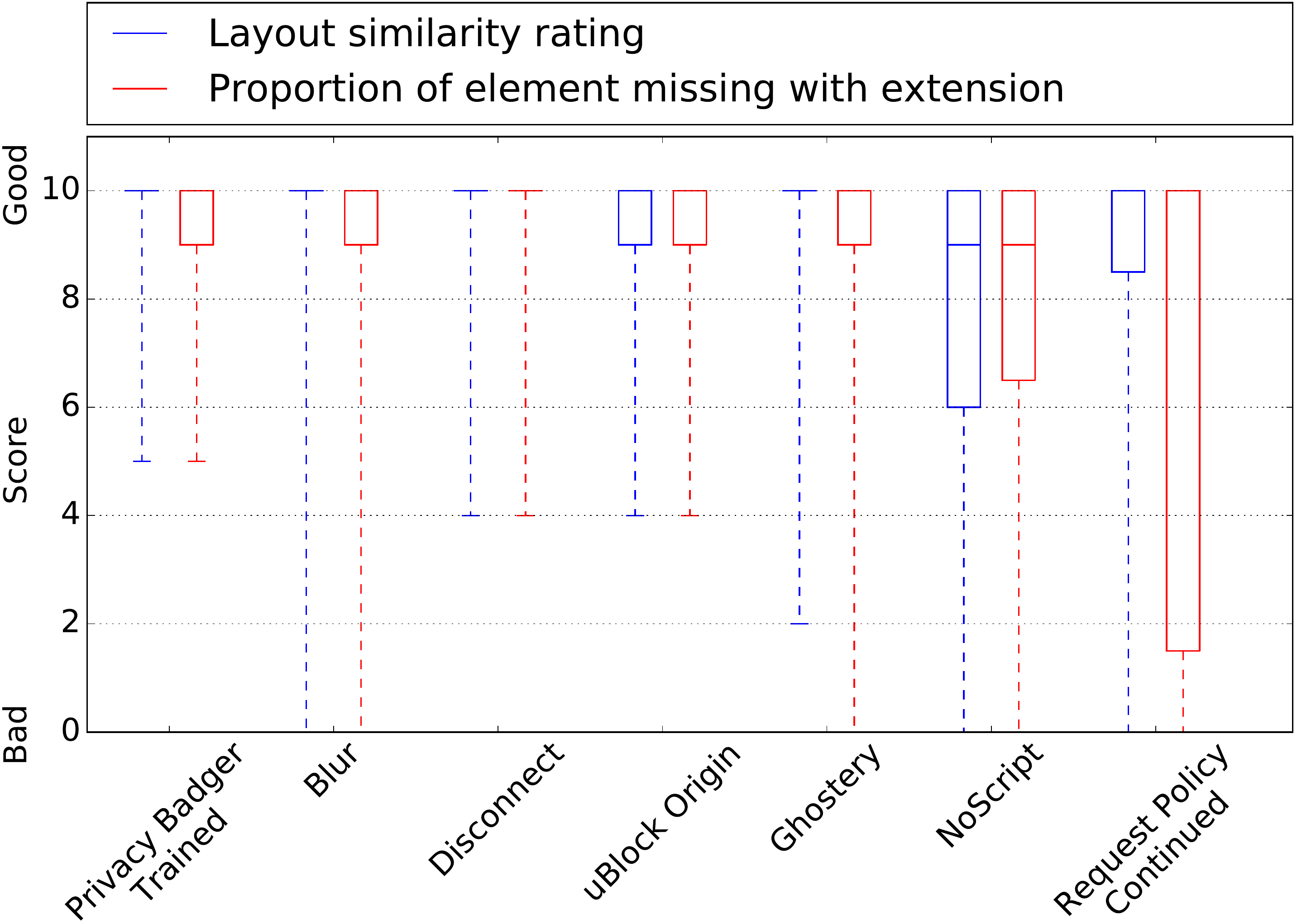}
  \caption{Webpage quality manual analysis. Users are presented two screenshots obtained with Firefox alone and with a privacy protection technique. The first question (blue) asks user to rate layout similarity. The second question (red) corresponds to the proportion of elements observed with Firefox alone also present when a privacy protection technique is used. Screenshots are captured on the Alexa Top 50. Boxplots extend to the upper and lower quartile and contain a line that represents the median. Dashed lines show min/max range. When not visible, median is located on top of the upper quartile.}
  \label{fig:manual_analysis}
\end{figure}

\subsubsection{Manual analysis}

The previous subsubsection analyzes webpages quality in terms of HTML and associated data.
This automated approach does not necessarily reflect the quality of the rendered webpage.
We perform a manual analysis to cope with this limitation.
We gather screenshots of the Alexa Top 50 without pages from the same entity but with distinct localization (e.g. Google and Amazon).
These screenshots were gathered in November 2016.
We use the seven techniques that provided the best privacy protection (see \autoref{sec:comparison_simple_metrics}): Privacy Badger trained, Blur, Disconnect, uBlock Origin, Ghostery, NoScript and RequestPolicy Continued.
For each webpage and protection technique, we display a picture that contains two screenshots side-to-side to the user: the left one captured with Firefox alone, the other with Firefox used with an extension.
We use the questions about layout similarity and missing elements presented in \autoref{sec:manual_analysis}.
\autoref{fig:manual_analysis} exposes our results.
The blue (resp. red) boxplot corresponds to the first (resp. second) question.
Overall, Privacy Badger, Blur, Disconnect, uBlockOrigin and Ghostery have a very small impact on rendered webpages.
On the other hand, RequestPolicy Continued and NoScript significantly impact on pages both in terms of layout and missing elements.
This is consistent with \autoref{sec:comparison_html_metrics} and previous results \cite{Krishnamurthy2007Measuring,Malandrino2013PrivacyLeakage}.
The manual analysis uncovers RequestPolicy Continued's strong impact on webpage quality which was not exposed by HTML-based metrics.

\begin{figure*}[ht!]
  \centering
  \subfloat[HTML-based vs manual analysis quality]{
    \includegraphics[width=0.33\textwidth]{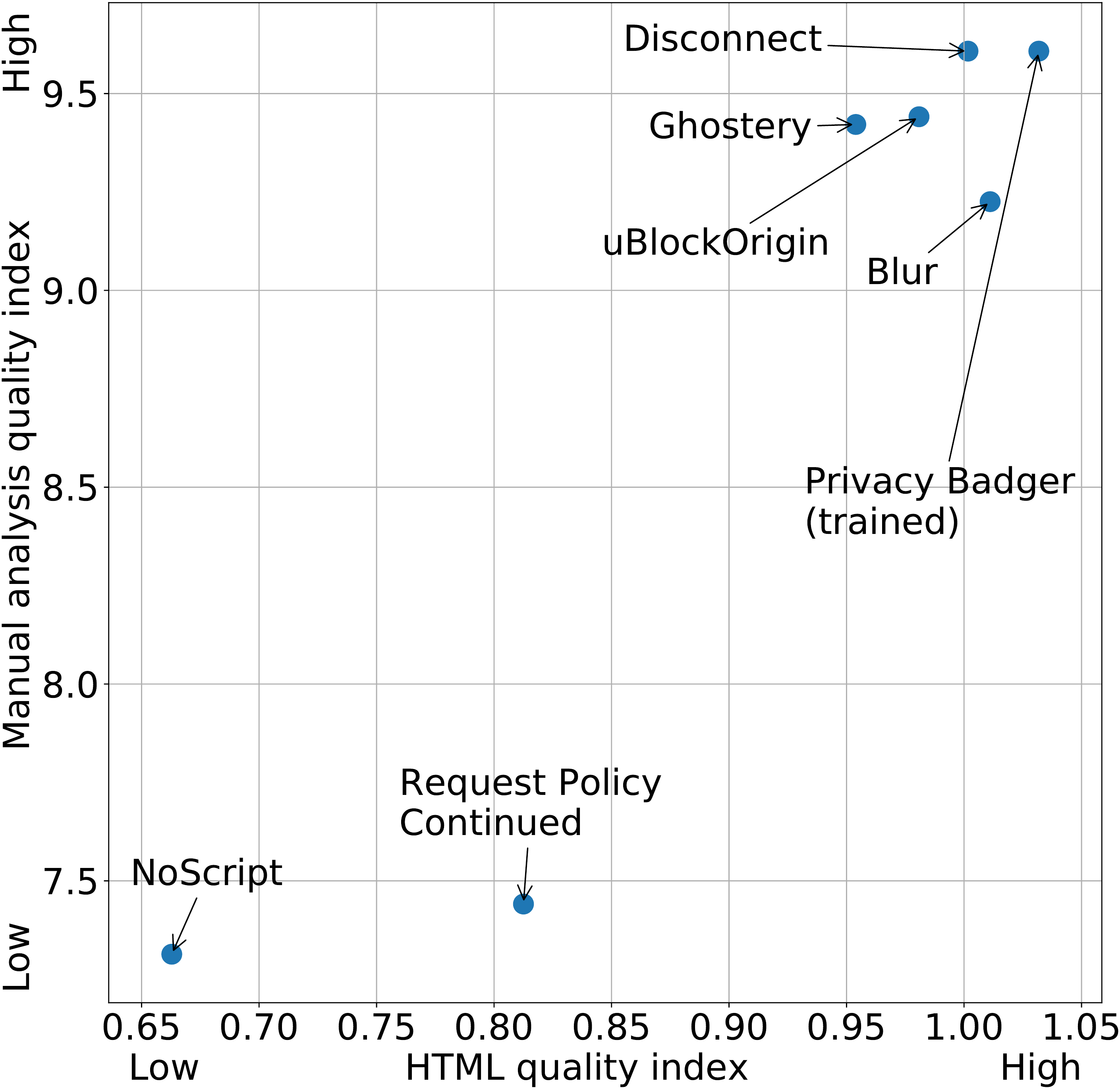}
    \label{fig:comparison_wqm_manual_analysis}
  }
  \subfloat[Privacy protection vs HTML-based quality]{
    \includegraphics[width=0.33\textwidth]{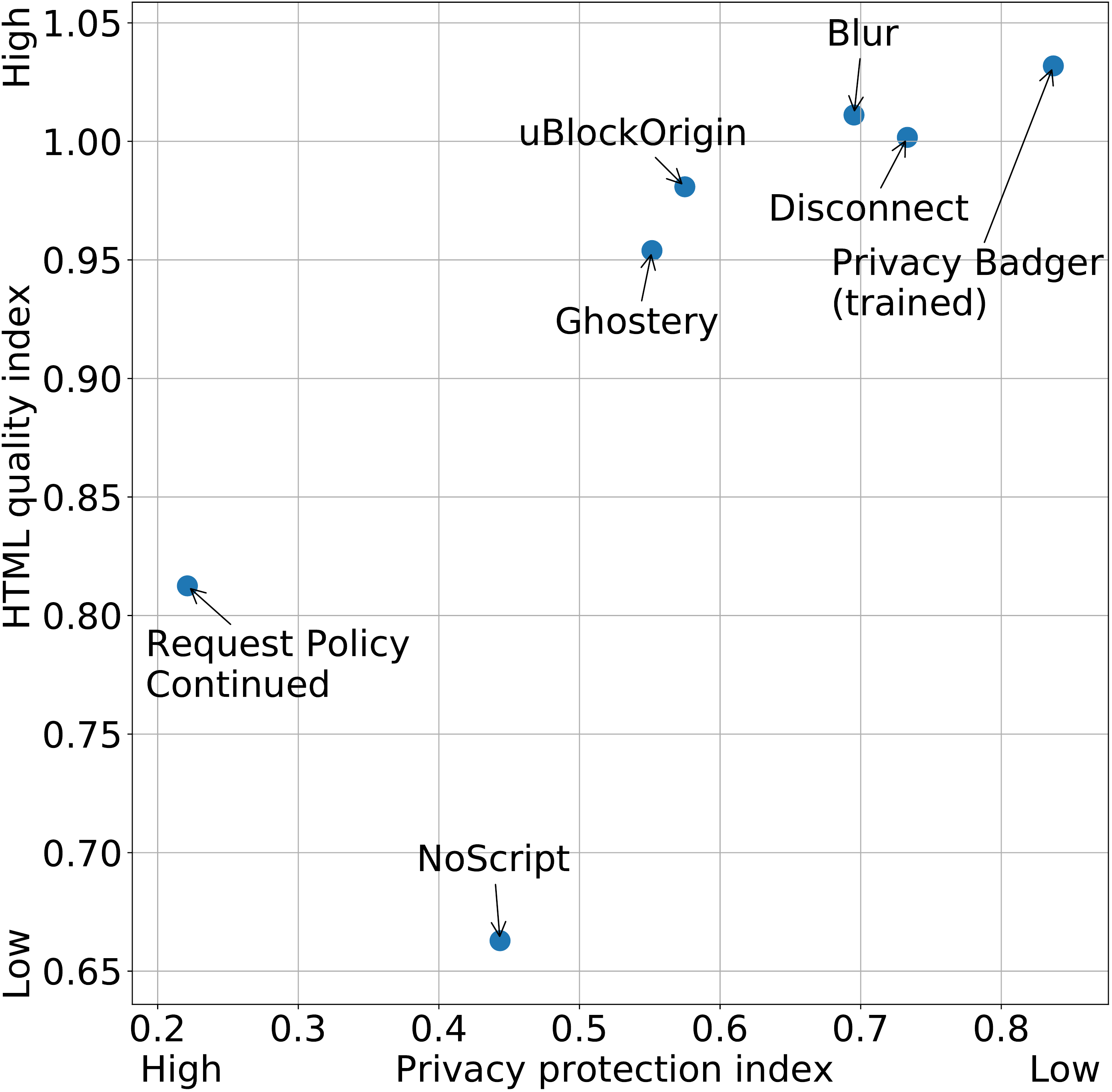}
    \label{fig:comparison_sm_wqm}
  }
  \subfloat[Privacy protection vs manual analysis]{
    \includegraphics[width=0.33\textwidth]{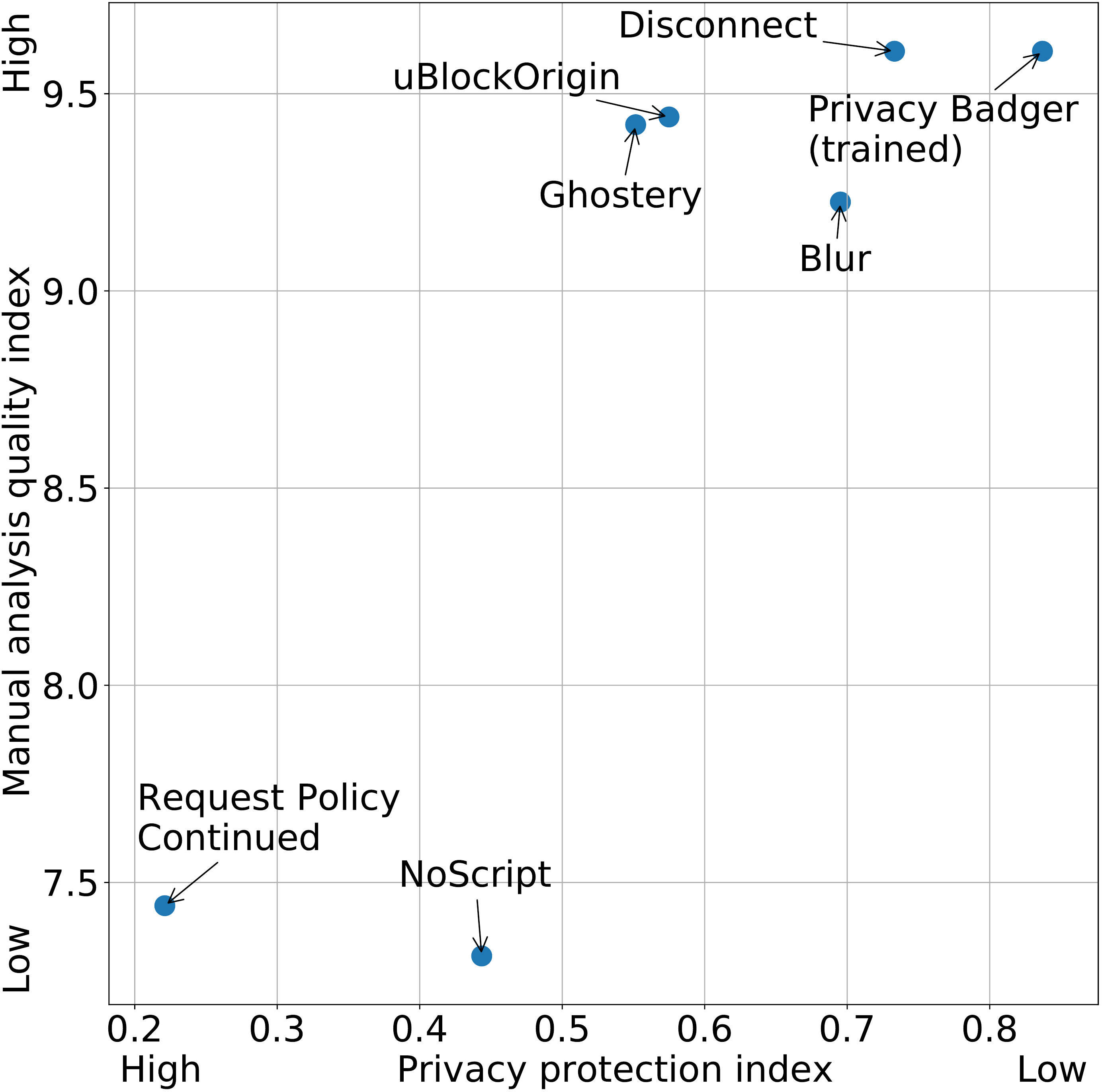}
    \label{fig:comparison_sm_manual_analysis}
  }

  \caption{Scatter plots of synthetic index for privacy protection, HTML-based quality and manual analysis-based quality.}
  \label{fig:synthetic}
\end{figure*}

\subsubsection*{Summary}

Both studies show that RequestPolicyContinued and NoScript reduce webpage quality.
Other extensions do not have such impact.

\subsection{Summary}

\autoref{fig:synthetic} summarizes our findings.
Each metric-based synthetic index is build as the mean of previously used metrics (see \autoref{sec:browsing_metrics} \autoref{sec:html_metrics}) normalized using the metric value obtained with Firefox used alone.
Manual analysis-based index is the mean of the mean rating of both questions.
Figure~\autoref{fig:comparison_wqm_manual_analysis} shows that our manual analysis captures the negative effects of \rpc that are not visible with HTML-based metrics.
Figures~\autoref{fig:comparison_sm_wqm} and \autoref{fig:comparison_sm_manual_analysis} demonstrate \rpc and NoScript show the best protection performances but impact browsing experience.
Ghostery and uBlock Origin protect users slightly less but have a smaller impact on webpage quality.
Remaining techniques provide average protection.

\section{Discussions}
\label{sec:discussions}

\subsection{Recommendations}

Our analysis shows that \rpc and Noscript are the most effective privacy protection techniques.
Similarly to previous work \cite{Krishnamurthy2007Measuring,Malandrino2013PrivacyLeakage}, we confirm that they affect webpage quality.
Users that want to avoid website breakage can instead use uBlock Origin or Ghostery.
The latter however needs to be configured which is difficult for users \cite{Leon2012Why}.
We note that previous work \cite{Merzdovnik2017Block,Traverso2017Benchmark} also recommended \ubo and Ghostery.
Their results were however noisy (percentiles are overlapping in \cite{Traverso2017Benchmark} and some techniques actually observed more fingerprintting than the default browser in \cite {Merzdovnik2017Block}) due to the use of single crawl to each website.
Our robust measurement methodology does not suffer from the same weakness, and show that uBlock Origin or Ghostery exhibit statistically better performance than other techniques.

Previous network traffic monitoring studies \cite{Metwalley2015The,Pujol2015Annoyed} show that users are more preoccupied by blocking ads than tracking.
Malloy et al. \cite{Malloy2016Ad} show that ad-blockers usage vary between 18\% and 37\% in analyzed countries (US, UK, Germany, France, and Canada).
Metwalley et al. show \cite{Metwalley2015The} that 10 to 18\% of users had installed AdBlock Plus while less than 3.5\% (resp. 2\%) were using DoNotTrackMe/Blur (resp. Ghostery).
Similarly, Pujol et al. \cite{Pujol2015Annoyed} speculate that out of the 19.7\% of users that install AdBlock Plus, less than 15\% of them setup the EasyPrivacy list.
We show that the default setup of AdBlock Plus has average performances but that it can be configured to achieve good results.
This task is however difficult for users \cite{Leon2012Why}.
We thus advise ad-blocking users that want to improve their privacy to directly change to more efficient extensions (see above).

Extensions that do not rely on manually maintained filters, Privacy Badger and MyTrackingChoices, exhibit average performances.
It however is not clear how their performances may be improved.
Analyzing extensions' behavior against existing blocking lists would help to improve their results.
Blocking list-based protection techniques are efficient but their scalability is limited due to human intervention.
Current efforts \cite{Bau2013Promising,Metwalley2015Unsupervised,Wu2015Tracker,Gugelmann2015Automated,Ikram2017Towards} to automatically build blocking list are thus extremely relevant.

Ghostery and uBlock Origin block a significant amount of resources that are not blocked by other protection techniques.
Both should thus be considered as relevant when assessing users' ability to protect themselves.

\subsection{Limitations}

In this work, we estimate the impact of privacy protection techniques on webpage quality in terms of missing elements and data (using HTML-based metrics \autoref{sec:html_metrics} and alteration to the webpage layout and elements (using a user manual analysis  \autoref{sec:manual_analysis}).
We however do not address the functionality loss.
We intend to explore this aspect in future work along two axes.
We plan to explore each website beyond its homepage to improve website coverage.

Several work \cite{Roesner2012Detecting,Bau2013Promising,Gugelmann2015Automated,Metwalley2015Unsupervised,Wu2015Tracker,Ikram2017Towards} automatically build blocking lists.
We did not evaluate them because produced lists are not publicly available.

Evaluated approaches aim at blocking tracking.
One side effect is advertisement blocking.
These approaches thus threaten the current economic model of many publishers on the Internet.
Techniques examined in this work however do not have the same impact on third party tracking.
For example, Ghostery and MyTrackingChoices can block tracking for specific website categories.
Similarly, Firefox's third party cookie blocking only partially impacts advertisement techniques (such as Real-Time Bidding, RTB) that rely on user interests.
Olejnik et al. \cite{Olejnik2014Selling} show that new users (i.e. users without any browsing history) are less valued by advertisers.
A user that blocks all third parties cookies thus reduces his customer value for advertisers, and, consequently, publisher's revenue.
This however does not completely block tracking and thus, advertisement transactions (e.g. bidding in RTB).
Indiscriminate third party blocking techniques (such as \rpc \cite{RequestPolicyContinued}) however have a much bigger impact on advertising.
They for example block interactions between ad exchange and bidders in the case of RTB.
All privacy protection techniques presented in this work may thus have very different impact on advertising. We intend to address this aspect in future work.

Ghostery \cite{Ghostery} and MyTrackingChoices \cite{Achara2016My} can block tracking for specific website categories, and thus, allow monetization for others.
This is obviously less aggressive than most approaches addressed in this work.
Achara et al. \cite{Achara2016My} however report that 30\% of users block the tracking for all categories.
These users thus have no reason to use these two techniques instead of other ones.

\subsection{Future work}

As shown in \autoref{sec:related_work}, existing work uses a vast array of metrics.
Some of them are tailored to very specific use cases (such as behavioral advertising \cite{Balebako2012Measuring,Carrascosa2015Always} and cookie-based mass-surveillance \cite{Englehardt2015Cookies}).
Similarly, data-related metrics present an obvious interest in a mobile context.
We intend to add new metrics to this work.
We also want to analyze the impact of all these protection techniques on specific attacks (browser fingerprinting \cite{Eckersley2010How,Englehardt2016Online}, cookie syncing \cite{Olejnik2014Selling,Acar2014The,Falahrastegar2016Tracking,Englehardt2016Online}).

Analyzing extensions inside other browsers, such as Chrome, Internet Explorer or Opera, would provide a wider picture of privacy protection techniques in the wild.
Selenium, the browser automation framework used by OpenWPM, supports these browsers.
OpenWPM, however, currently only supports Firefox.
Furthermore, browsers include more and more privacy protection (e.g. Firefox \cite{Kontaxis2015Tracking}, Brave \cite{Brave}, Cliqz \cite{Yu2016Tracking}).
Cliqz and Brave are not supported by Selenium, which make both browsers impossible to use with OpenWPM (see \autoref{sec:data_collection}).
We however intend to add them later on.

While some works use domain name-based heuristics \cite{Falahrastegar2014Rise,Olejnik2014Selling} to identify third parties, others use blocking lists-based extensions as ground-truth (see \autoref{table:related_work_techniques}).
When the latter method is used, tracking's long tail \cite{Englehardt2016Online} may cause many false negatives.
Beyond our analysis in \autoref{subsec:extensions}, we intend to investigate blocking lists more thoroughly.

\section{Conclusions}

This work extensively compares existing privacy protection techniques against third party tracking and provides four contributions.
First, we propose a robust methodology to compare privacy protection techniques when crawling many websites, and quantify measurement error.
We use the privacy footprint \cite{Krishnamurthy2006Generating} and apply the Kolmogorov-Smirnov (KS) test on browsing metrics to evaluate privacy protection.
This test is likewise applied to HTML-based metrics to assess the impact on webpage quality.
We also design a manual analysis to complement HTML-based metrics.
Our second contribution is a privacy protection techniques comparison in terms of overlap and performances.
We show that the most popular privacy protection techniques exhibit a common blocking baseline.
We also highlight the fact that some extensions, namely Ghostery and uBlock Origin, have a very specific behavior and are the only ones to block some domains.
Our third contribution is a comparison of the impact of privacy protection techniques on webpage quality.
Our fourth contribution is a set of usage recommendations for end-users, and research recommendation for the scientific community.
Ghostery and uBlock Origin provide the best trade-off between protection and webpage quality.
Ghostery however requires a configuration step which is difficult for users \cite{Leon2012Why}.
The \rpc and NoScript extensions exhibit the best performances but reduce webpage quality.
Ghostery and uBlock Origin use manually built blocking lists which are cumbersome to maintain.
Research efforts should focus on improving existing approaches that do not rely on blocking lists (such as Privacy Badger or MyTrackingChoices) and automating blocking list building.

\bibliographystyle{IEEEtran}
\bibliography{references}

\begin{thebibliography}{10}
\providecommand{\url}[1]{#1}
\csname url@samestyle\endcsname
\providecommand{\newblock}{\relax}
\providecommand{\bibinfo}[2]{#2}
\providecommand{\BIBentrySTDinterwordspacing}{\spaceskip=0pt\relax}
\providecommand{\BIBentryALTinterwordstretchfactor}{4}
\providecommand{\BIBentryALTinterwordspacing}{\spaceskip=\fontdimen2\font plus
\BIBentryALTinterwordstretchfactor\fontdimen3\font minus
  \fontdimen4\font\relax}
\providecommand{\BIBforeignlanguage}[2]{{%
\expandafter\ifx\csname l@#1\endcsname\relax
\typeout{** WARNING: IEEEtran.bst: No hyphenation pattern has been}%
\typeout{** loaded for the language `#1'. Using the pattern for}%
\typeout{** the default language instead.}%
\else
\language=\csname l@#1\endcsname
\fi
#2}}
\providecommand{\BIBdecl}{\relax}
\BIBdecl

\bibitem{Krishnamurthy2006Generating}
B.~Krishnamurthy and C.~E. Wills, ``Generating a privacy footprint on the
  internet,'' in \emph{IMC 2006}, pp. 65--70.

\bibitem{Acar2014The}
G.~Acar, C.~Eubank, S.~Englehardt, M.~Juarez, A.~Narayanan, and C.~Diaz, ``The
  web never forgets: Persistent tracking mechanisms in the wild,'' in \emph{CCS
  2014}, pp. 674--689.

\bibitem{Soltani2010Flash}
A.~Soltani, S.~Canty, Q.~Mayo, L.~Thomas, and C.~J. Hoofnagle, ``Flash cookies
  and privacy,'' in \emph{SSRN}, 2009, pp. 158--163.

\bibitem{Eckersley2010How}
P.~Eckersley, ``How unique is your web browser?'' in \emph{PETS 2010}, pp.
  1--18.

\bibitem{DoNotTrack}
\BIBentryALTinterwordspacing
``{D}o{N}ot{T}rack,'' accessed: 2017-07-30. [Online]. Available:
  \url{https://tools.ietf.org/html/draft-mayer-do-not-track-00}
\BIBentrySTDinterwordspacing

\bibitem{Krishnamurthy2007Measuring}
B.~Krishnamurthy, D.~Malandrino, and C.~E. Wills, ``Measuring privacy loss and
  the impact of privacy protection in web browsing,'' in \emph{SOUPS 2007}, pp.
  52--63.

\bibitem{Mayer2012Third}
J.~R. Mayer and J.~C. Mitchell, ``Third-party web tracking: Policy and
  technology,'' in \emph{SP 2012}, pp. 413--427.

\bibitem{Balebako2012Measuring}
R.~Balebako, P.~Leon, R.~Shay, B.~Ur, Y.~Wang, and L.~Cranor, ``Measuring the
  effectiveness of privacy tools for limiting behavioral advertising,'' in
  \emph{W2SP 2012}.

\bibitem{Hill2014Comparative}
\BIBentryALTinterwordspacing
R.~Hill, ``Comparative benchmarks against widely used blockers: Top 15 most
  popular news websites,'' 2014, accessed: 2017-07-30. [Online]. Available:
  \url{https://github.com/gorhill/httpswitchboard/wiki/Comparative-benchmarks-against-widely-used-blockers:-Top-15-Most-Popular-News-Websites}
\BIBentrySTDinterwordspacing

\bibitem{Hill2015uBlock}
\BIBentryALTinterwordspacing
------, ``ublock and others: Blocking ads, trackers, malwares,'' 2015,
  accessed: 2017-07-30. [Online]. Available:
  \url{https://github.com/gorhill/uBlock/wiki/uBlock-and-others%3A-Blocking-ads%2C-trackers%2C-malwares}
\BIBentrySTDinterwordspacing

\bibitem{Wills2016What}
C.~E. Wills and D.~C. Uzunoglu, ``What ad blockers are (and are not) doing,''
  in \emph{HotWeb 2016}, pp. 72--77.

\bibitem{Merzdovnik2017Block}
G.~Merzdovnik, M.~Huber, D.~Buhov, N.~Nikiforakis, S.~Neuner, M.~Schmiedecker,
  and E.~Weippl, ``Block me if you can: A large-scale study of tracker-blocking
  tools,'' in \emph{EuroSP 2017}.

\bibitem{Traverso2017Benchmark}
S.~Traverso, M.~Trevisan, L.~Giannantoni, M.~Mellia†, and H.~Metwalley,
  ``Benchmark and comparison of tracker-blockers:should you trust them?'' in
  \emph{TMA 2017}.

\bibitem{Bujlow2017Web}
T.~Bujlow, V.~Carela-Espa{\~n}ol, J.~Sol{\'e}-Pareta, and P.~Barlet-Ros, ``A
  survey on web tracking: Mechanisms, implications, and defenses,''
  \emph{Proceedings of the IEEE}, vol.~PP, no.~99, pp. 1--35, 2017.

\bibitem{Estrada2016Online}
J.~Estrada-Jim{\'e}nez, J.~Parra-Arnau, A.~Rodr{\'\i}guez-Hoyos, and
  J.~Forn{\'e}, ``Online advertising: Analysis of privacy threats and
  protection approaches,'' \emph{Computer Communications}, 2016.

\bibitem{AdBlockPlus}
\BIBentryALTinterwordspacing
``{A}d{B}lock {P}lus,'' accessed: 2017-07-30. [Online]. Available:
  \url{https://adblockplus.org/}
\BIBentrySTDinterwordspacing

\bibitem{uBlockOrigin}
\BIBentryALTinterwordspacing
``u{B}lock {O}rigin,'' accessed: 2017-07-30. [Online]. Available:
  \url{https://github.com/gorhill/uBlock}
\BIBentrySTDinterwordspacing

\bibitem{Ghostery}
\BIBentryALTinterwordspacing
``Ghostery,'' accessed: 2017-07-30. [Online]. Available:
  \url{https://www.ghostery.com/}
\BIBentrySTDinterwordspacing

\bibitem{Disconnect}
\BIBentryALTinterwordspacing
``Disconnect,'' accessed: 2017-07-30. [Online]. Available:
  \url{https://disconnect.me/}
\BIBentrySTDinterwordspacing

\bibitem{NoTrace}
\BIBentryALTinterwordspacing
``{N}o{T}race,'' accessed: 2017-07-30. [Online]. Available:
  \url{http://www.isislab.it/projects/NoTrace/}
\BIBentrySTDinterwordspacing

\bibitem{Blur}
\BIBentryALTinterwordspacing
``Blur,'' accessed: 2017-07-30. [Online]. Available:
  \url{https://dnt.abine.com}
\BIBentrySTDinterwordspacing

\bibitem{Achara2016My}
J.~P. Achara, J.~Parra-Arnau, and C.~Castelluccia, ``Mytrackingchoices:
  Pacifying the ad-block war by enforcing user privacy preferences,'' in
  \emph{WEIS 2016}.

\bibitem{PrivacyBadger}
\BIBentryALTinterwordspacing
``{P}rivacy {B}adger,'' accessed: 2017-07-30. [Online]. Available:
  \url{https://www.eff.org/fr/privacybadger}
\BIBentrySTDinterwordspacing

\bibitem{NoScript}
\BIBentryALTinterwordspacing
``{N}o{S}cript,'' accessed: 2017-07-30. [Online]. Available:
  \url{https://noscript.net/}
\BIBentrySTDinterwordspacing

\bibitem{RequestPolicyContinued}
\BIBentryALTinterwordspacing
``{R}equest{P}olicy{C}ontinued,'' accessed: 2017-07-30. [Online]. Available:
  \url{https://requestpolicycontinued.github.io/}
\BIBentrySTDinterwordspacing

\bibitem{HTTPSEverywhere}
\BIBentryALTinterwordspacing
``{HTTPS} {E}verywhere,'' accessed: 2017-07-30. [Online]. Available:
  \url{https://www.eff.org/fr/https-everywhere}
\BIBentrySTDinterwordspacing

\bibitem{Decentraleyes}
\BIBentryALTinterwordspacing
``Decentraleyes,'' accessed: 2017-07-30. [Online]. Available:
  \url{https://decentraleyes.org}
\BIBentrySTDinterwordspacing

\bibitem{WebOfTrust}
\BIBentryALTinterwordspacing
``{W}eb{O}f{T}rust,'' accessed: 2017-07-30. [Online]. Available:
  \url{https://www.mywot.com/}
\BIBentrySTDinterwordspacing

\bibitem{Nikiforakis2013Cookieless}
N.~Nikiforakis, A.~Kapravelos, W.~Joosen, C.~Kruegel, F.~Piessens, and
  G.~Vigna, ``Cookieless monster: Exploring the ecosystem of web-based device
  fingerprinting,'' in \emph{SP 2013}, pp. 541--555.

\bibitem{Vincent2017Ghostery}
\BIBentryALTinterwordspacing
J.~Vincent, ``Ghostery has been bought by the developer of a privacy-focused
  browser,'' 2 2017, accessed: 2017-07-30. [Online]. Available:
  \url{http://www.theverge.com/2017/2/15/14622484/ghostery-ad-tracking-plug-in-cliqz}
\BIBentrySTDinterwordspacing

\bibitem{NAI}
\BIBentryALTinterwordspacing
``Network advertising industry,'' accessed: 2017-07-30. [Online]. Available:
  \url{http://www.networkadvertising.org/choices/}
\BIBentrySTDinterwordspacing

\bibitem{DAA}
\BIBentryALTinterwordspacing
``Digital advertising alliance,'' accessed: 2017-07-30. [Online]. Available:
  \url{http://www.aboutads.info/choices/}
\BIBentrySTDinterwordspacing

\bibitem{BeefTaco}
\BIBentryALTinterwordspacing
``Beeftaco,'' accessed: 2017-07-30. [Online]. Available:
  \url{https://jmhobbs.github.io/beef-taco/}
\BIBentrySTDinterwordspacing

\bibitem{AcceptableAds}
\BIBentryALTinterwordspacing
``Allowing acceptable ads in adblock plus - agreements,'' accessed: 2017-07-30.
  [Online]. Available: \url{https://adblockplus.org/acceptable-ads-agreements}
\BIBentrySTDinterwordspacing

\bibitem{EasyList_EasyPrivacy}
\BIBentryALTinterwordspacing
``{E}asy{List}/{E}asy{P}rivacy,'' accessed: 2017-07-30. [Online]. Available:
  \url{https://easylist.to/}
\BIBentrySTDinterwordspacing

\bibitem{Eckert2016Web}
\BIBentryALTinterwordspacing
V.~S. Eckert, J.~Klofta, and J.~L. Strozyk, ````{W}eb of {T}rust'' sp\"{a}ht
  nutzer aus,'' 11 2016, accessed: 2017-07-30. [Online]. Available:
  \url{https://www.tagesschau.de/inland/tracker-online-103.html}
\BIBentrySTDinterwordspacing

\bibitem{Olejnik2017Proposed}
\BIBentryALTinterwordspacing
L.~Olejnik, ``Proposed amendments to eprivacy regulation are great,'' 6 2017,
  accessed: 2017-07-30. [Online]. Available:
  \url{https://blog.lukaszolejnik.com/proposed-amendments-to-eprivacy-regulation-are-great}
\BIBentrySTDinterwordspacing

\bibitem{Firefox}
\BIBentryALTinterwordspacing
``Firefox,'' accessed: 2017-07-30. [Online]. Available:
  \url{https://www.mozilla.org/en-US/firefox}
\BIBentrySTDinterwordspacing

\bibitem{Kontaxis2015Tracking}
G.~Kontaxis and M.~Chew, ``Tracking protection in firefox for privacy and
  performance,'' in \emph{W2SP 2015}.

\bibitem{Brave}
\BIBentryALTinterwordspacing
``Brave,'' accessed: 2017-07-30. [Online]. Available: \url{https://brave.com/}
\BIBentrySTDinterwordspacing

\bibitem{CLiqz}
\BIBentryALTinterwordspacing
``Cliqz,'' accessed: 2017-07-30. [Online]. Available:
  \url{https://cliqz.com/en}
\BIBentrySTDinterwordspacing

\bibitem{AdBlock}
\BIBentryALTinterwordspacing
``{A}d{B}lock,'' accessed: 2017-07-30. [Online]. Available:
  \url{https://getadblock.com/}
\BIBentrySTDinterwordspacing

\bibitem{Castelluccia2013Data}
C.~Castelluccia, S.~Grumbach, and L.~Olejnik, ``Data harvesting 2.0: from the
  visible to the invisible web,'' in \emph{WEIS 2013}.

\bibitem{Fruchter2015Variations}
N.~Fruchter, H.~Miao, S.~Stevenson, and R.~Balebako, ``Variations in tracking
  in relation to geographic location,'' in \emph{W2SP 2015}.

\bibitem{Carrascosa2015Always}
J.~M. Carrascosa, J.~Mikians, R.~Cuevas, V.~Erramilli, and N.~Laoutaris, ``I
  always feel like somebody's watching me. measuring online behavioural
  advertising,'' in \emph{CoNext 2015}.

\bibitem{Metwalley2015The}
H.~Metwalley, S.~Traverso, M.~Mellia, S.~Miskovic, and M.~Baldi, ``The online
  tracking horde: a view from passive measurements,'' in \emph{TMA 2015}, pp.
  111--125.

\bibitem{Englehardt2015Cookies}
S.~Englehardt, D.~Reisman, C.~Eubank, P.~Zimmerman, J.~Mayer, A.~Narayanan, and
  E.~W. Felten, ``Cookies that give you away: The surveillance implications of
  web tracking,'' in \emph{WWW 2015}, pp. 289--299.

\bibitem{Englehardt2016Online}
S.~Englehardt and A.~Narayanan, ``Online tracking: A 1-million-site measurement
  and analysis,'' in \emph{CCS 2016}.

\bibitem{Malandrino2013PrivacyAwareness}
D.~Malandrino, A.~Petta, V.~Scarano, L.~Serra, R.~Spinelli, and
  B.~Krishnamurthy, ``Privacy awareness about information leakage: Who knows
  what about me?'' in \emph{WPES 2013}, pp. 279--284.

\bibitem{Malandrino2013PrivacyLeakage}
D.~Malandrino and V.~Scarano, ``Privacy leakage on the web: Diffusion and
  countermeasures,'' \emph{Computer Networks}, vol.~57, no.~14, pp. 2833--2855,
  2013.

\bibitem{Gugelmann2015Automated}
D.~Gugelmann, M.~Happe, B.~Ager, and V.~Lenders, ``An automated approach for
  complementing ad blockers’ blacklists,'' in \emph{PETS 2015}, pp. 282--298.

\bibitem{Metwalley2015Unsupervised}
H.~Metwalley, S.~Traverso, and M.~Mellia, ``Unsupervised detection of web
  trackers,'' in \emph{GLOBECOM 2015}, pp. 1--6.

\bibitem{Wu2015Tracker}
Q.~Wu, Q.~Liu, Y.~Zhang, and G.~Wen, ``Trackerdetector: A system to detect
  third-party trackers through machine learning,'' \emph{Computer Networks},
  vol.~91, pp. 164--173, 2015.

\bibitem{Yu2016Tracking}
Z.~Yu, S.~Macbeth, K.~Modi, and J.~M. Pujol, ``Tracking the trackers,'' in
  \emph{WWW 2016}, pp. 121--132.

\bibitem{Ikram2017Towards}
M.~Ikram, H.~J. Asghar, M.~A. Kaafar, A.~Mahanti, and B.~Krishnamurthy,
  ``Towards seamless tracking-free web: Improved detection of trackers via
  one-class learning,'' ser. PETS 2017.

\bibitem{Leon2012Why}
P.~Leon, B.~Ur, R.~Shay, Y.~Wang, R.~Balebako, and L.~Cranor, ``Why johnny
  can't opt out: A usability evaluation of tools to limit online behavioral
  advertising,'' in \emph{CHI 2012}, pp. 589--598.

\bibitem{Malandrino2013How}
D.~Malandrino, V.~Scarano, and R.~Spinelli, ``How increased awareness can
  impact attitudes and behaviors toward online privacy protection,'' in
  \emph{SocialCom 2013}, pp. 57--62.

\bibitem{Krishnamurthy2009Privacy}
B.~Krishnamurthy and C.~Wills, ``Privacy diffusion on the web: a longitudinal
  perspective,'' in \emph{WWW 2009}, pp. 541--550.

\bibitem{Roesner2012Detecting}
F.~Roesner, T.~Kohno, and D.~Wetherall, ``Detecting and defending against
  third-party tracking on the web,'' in \emph{NSDI 2012}, pp. 12--26.

\bibitem{Fifield2015Fingerprinting}
D.~Fifield and S.~Egelman, ``Fingerprinting web users through font metrics,''
  in \emph{FC 2015}, pp. 107--124.

\bibitem{Olejnik2015Leaking}
{\L}.~Olejnik, G.~Acar, C.~Castelluccia, and C.~Diaz, ``The leaking battery,''
  in \emph{DPM 2015}, pp. 254--263.

\bibitem{Mowery2012Pixel}
K.~Mowery and H.~Shacham, ``Pixel perfect: Fingerprinting canvas in html5,'' in
  \emph{W2SP 2012}.

\bibitem{Cao2017Cross}
Y.~Cao, S.~Li, and E.~Wijmans, ``({C}ross-){B}rowser {F}ingerprinting via {OS}
  and {H}ardware {L}evel {F}eatures,'' in \emph{NDSS 2017}.

\bibitem{Acar2013FP}
G.~Acar, M.~Juarez, N.~Nikiforakis, C.~Diaz, S.~G{\"u}rses, F.~Piessens, and
  B.~Preneel, ``Fpdetective: dusting the web for fingerprinters,'' in \emph{CCS
  2013}, pp. 1129--1140.

\bibitem{Lerner2016Internet}
A.~Lerner, A.~K. Simpson, T.~Kohno, and F.~Roesner, ``Internet jones and the
  raiders of the lost trackers: An archaeological study of web tracking from
  1996 to 2016,'' in \emph{Usenix Security 2016}, pp. 997--1013.

\bibitem{Ghosh2011Selling}
A.~Ghosh and A.~Roth, ``Selling privacy at auction,'' in \emph{EC 2011}, pp.
  199--208.

\bibitem{Olejnik2014Selling}
L.~Olejnik, T.~Minh-Dung, and C.~Castelluccia, ``Selling off privacy at
  auction,'' in \emph{NDSS 2014}, pp. 104--114.

\bibitem{Bau2013Promising}
J.~Bau, J.~Mayer, H.~Paskov, and J.~C. Mitchell, ``A promising direction for
  web tracking countermeasures,'' in \emph{W2SP 2013}.

\bibitem{Kalavri2016Like}
V.~Kalavri, J.~Blackburn, M.~Varvello, and K.~Papagiannaki, ``Like a pack of
  wolves: Community structure of web trackers,'' pp. 42--54.

\bibitem{Alexa}
\BIBentryALTinterwordspacing
``Alexa top 500 sites on the web,'' accessed: 2017-07-30. [Online]. Available:
  \url{http://www.alexa.com}
\BIBentrySTDinterwordspacing

\bibitem{PublicSuffixList}
\BIBentryALTinterwordspacing
``Public suffix list,'' accessed: 2017-07-30. [Online]. Available:
  \url{https://publicsuffix.org/}
\BIBentrySTDinterwordspacing

\bibitem{Guha2010Challenges}
S.~Guha, B.~Cheng, and P.~Francis, ``Challenges in measuring online advertising
  systems,'' in \emph{IMC 2010}, pp. 81--87.

\bibitem{Pass2006Picture}
G.~Pass, A.~Chowdhury, and C.~Torgeson, ``A picture of search.''
  \emph{InfoScale}, vol. 152, p.~1, 2006.

\bibitem{Pujol2015Annoyed}
E.~Pujol, O.~Hohlfeld, and A.~Feldmann, ``Annoyed users: Ads and ad-block usage
  in the wild,'' in \emph{IMC 2016}, pp. 93--106.

\bibitem{Malloy2016Ad}
M.~Malloy, M.~McNamara, A.~Cahn, and P.~Barford, ``Ad blockers: Global
  prevalence and impact,'' in \emph{IMC 2016}, pp. 119--125.

\bibitem{Falahrastegar2016Tracking}
M.~Falahrastegar, H.~Haddadi, S.~Uhlig, and R.~Mortier, ``Tracking personal
  identifiers across the web,'' in \emph{PAM 2016}, pp. 30--41.

\bibitem{Falahrastegar2014Rise}
------, ``The rise of panopticons: Examining region-specific third-party web
  tracking.'' in \emph{TMA 2014}, pp. 104--114.

\end{thebibliography}

\end{document}